\documentclass[twocolumn, superscriptaddress, prmaterials, amsmath,amssymb,noeprint]{revtex4-2}

\usepackage{hyperref}
\hypersetup{backref,colorlinks=true,allcolors=prussianblue}
\usepackage{xr}
\usepackage{graphicx}
\usepackage[caption=false]{subfig}

\usepackage[table]{xcolor}
\definecolor{prussianblue}{rgb}{0.0, 0.19, 0.33}

\usepackage{setspace}
\usepackage{float}
\usepackage{indentfirst}

\usepackage{tcolorbox}
\usepackage{todonotes}

\def\fig#1{Fig.~\ref{fig:#1}}
\def\eq#1{Eq.~\eqref{eq:#1}}
\def\tab#1{Table~\ref{tab:#1}}

\def\der#1{d\mathbf{#1}\;}

\usepackage{tikz}
\usetikzlibrary{calc,arrows,shapes,positioning}
\usetikzlibrary{patterns,snakes}
\usetikzlibrary{positioning,decorations.pathreplacing}

\definecolor{sangria}{rgb}{0.57, 0.0, 0.04}
\definecolor{arsenic}{rgb}{0.23, 0.27, 0.29}
\definecolor{prussianblue}{rgb}{0.0, 0.19, 0.33}
\definecolor{phthalogreen}{rgb}{0.07, 0.21, 0.14}
\definecolor{dgreen}{rgb}{0.0, 0.4, 0.2}

\begin{document}

\author{Okan K. Orhan}
\author{Mohamed Hendy}
\author{Mauricio Ponga} \email[Corresponding author: ]{mponga@mech.ubc.ca}
\affiliation{Department of Mechanical Engineering, University of British Columbia, 2054 - 6250 Applied Science Lane, Vancouver, BC, V6T 1Z4, Canada}

\title{Electronic effects on the radiation damage in high-entropy alloys}

\begin{abstract}
High-entropy alloys (HEAs) are exceptional candidates for radiation-resistant materials due to their complex local chemical environment and slow defect migration. 
Despite commonly overlooked, electronic effects on defects evolution in radiation environments also play a crucial role by dissipating excess energy through electron-phonon coupling and electronic heat conduction during cascade events.  
We present a systematic study on electronic properties in random-solid solutions (RSS) in four and five principal elements HEAs and their effect on defect formation, clustering, and recombination. 
Electronic properties, including electron-phonon coupling factor ($G_\mathrm{e-ph}$), the electronic specific heat ($C_\mathrm{e}$), and the electronic thermal conductivity ($\kappa_\mathrm{e}$), are computed within first-principles calculations. 
Using the two-temperature molecular dynamics simulations, we show that the electron-phonon coupling factor and electronic specific heat play a critical role in Frenkel pairs formation. 
Specifically, the electron-phonon coupling factor quickly dissipates the kinetic energy during primary knock-on atom events via plasmon excitations and is subsequently dissipated via the free-electrons conduction. 
We show that these effects are more critical than the elastic distortion effects produced by the atomic mismatch. 
Of tremendous interest, we show that including lighter elements helps to increase $G_\mathrm{e-ph}$ suggesting the possibility to improve radiation resistance in HEA through optimal composition.
\end{abstract}

\maketitle

\section{Introduction}
High-entropy alloys (HEA) are {\color{black} \textit{mostly}} single-phased multi-principal-elements (MPE) alloys, often produced by  \textit{nearly} equimolar mixing of four, or more elements~\cite{CANTOR2004213,doi:10.1002/adem.200300567,Yeh2006}.
HEAs have been shown to exhibit exceptional thermal stability~\cite{Senkov2012}, hardness and strength ~\cite{CHENG20113185,GALI201374,CHEN201639,XIAN2017229,LI201835,CAI2019281},  wear resistance~\cite{Huang200474, Hsu2004},  and  oxidation resistance~\cite{Huang200474,KUMAR2017154}. 
In particular, the equimolar chromium-cobalt-iron-nickel HEA (CrFeCoNi) has attracted considerable interest as it can serve as a base template to  exploring non-equimolar compositions, or adding other elements such as manganese (Mn), copper (Cu), aluminum (Al), titanium (Ti), niobium (Nb), molybdenum (Mo) and many more to achieve exceptional materials properties~\cite{WU2014428,Gludovatz1153,WU2014131,
CHUANG20116308,ULLAH201617,Tong2005,ROJAS2022162309,10.3389/fmats.2021.816610}.

In high-radiation environments, highly energetic particle beams excite primary knock-on atoms (PKAs) out when they travel through solids, which in turn create self-interstitial atoms (SIA) and vacancies, called Frenkel pairs (FPs). 
Formation and clustering of FPs locally distort lattices, resulting in dislocation loops and void growth.
As these FPs recombine and migrate through the lattice, they generate significant microstructural changes over time, critically affecting the mechanical properties of materials subjected to radiation environments. 
Clustering of defected atoms commonly reduces the structural integrity of materials by causing swelling, creep, and irradiation-induced stress corrosion cracking~\cite{WAS2012195}. 

In the case of HEA, the complex local chemical environment and atomic-size mismatch create well-known elastic lattice distortions and core effects such as sluggish diffusion kinetics~\cite{Yeh2006}.
These effects help to reduce defect formation, slow down defect diffusion and accumulation, and promote vacancy-interstitial recombination, leading to lower radiation damage compared to pure metals and traditional alloys~\cite{TSAI20134887,Zhang2015}.
For instance, CrMnFeCoNi preserves its majority face-centered cubic phase even over $40$~DPA irradiation throughout the temperature range of $\sim 298-773$~K~\cite{NAGASE201532}. 
Thus, HEAs are promising candidates for structural materials exposed to high-radiation environments ~\cite{ULLAH201617,dada2019high} such as in nuclear reactors~\cite{Xia2015,ULLAH201617} and other environments such as space applications~\cite{dada2019high}.

Due to the many challenges that appear during the testing of materials under radiation environments, experimental studies are difficult to perform and time-consuming. 
At the same time, since the formation, migration, clustering, as well as recombination of defects occur within a few femtoseconds (fs) to nanoseconds (ns) of the primary cascade event, it is pretty challenging to \textit{in-situ} unravel mechanisms behind the defect evolution. 
Thus, the community heavily relies on numerical simulations to understand and predict radiation damage under radiation, especially for novel materials. 
In that sense, atomistic simulations based on either molecular dynamics (MD) or \emph{ab-initio} molecular dynamics provide a framework to study materials under radiation environments systematically \cite{ULLAH201617,DeLuigi:2021}. 
Being an atomistic technique, MD fully resolves the time and atomic environment and naturally describes the interactions between PKA and lattice, generating defects as the PKA losses its energy. 
These insights can then be incorporated in Monte-Carlo codes to predict damage in macroscopic specimens such as those used in engineering applications~\cite{Fluka:2015}. 
However, a drawback of classical MD is that it neglects electronic effects in materials. 

In metallic materials such as HEA, the excess energy due to high-radiation exposure is deposited into phonons and electrons, resulting in defect formation and microstructural changes. 
Remarkably, the primary heat carriers in metallic materials are free electrons, which result in thermal conductivities that are much higher than the lattice contribution~\cite{doi:10.1179/imtr.1986.31.1.197}.
However, in MPE alloys, these contributions have to be carefully analyzed since they might change depending on the composition and defect population~\cite{ziman_1972,makinson1938thermal}. 
Therefore, incorporating electronic effects in radiation damage is paramount for metallic materials to accurately understand and predict the damage. 

Due to the potential applications of MPE alloys as radiation-resistant materials, several studies have recently investigated their performance using a combination of experiments and numerical simulations, providing a fundamental understanding of the damage mechanisms during radiation in MPE alloys~\cite{KUMAR2016230,PhysRevLett.116.135504,Lu2016,BARR201880,DeLuigi:2021}. 
Remarkably, while the numerical simulations carried out in these studies include phonon effects in MPE alloys, electronic effects have been almost completely ignored in these works, making these results applicable -but also essentially limited- to small PKA energies.
In particular, electronic effects become more predominant at moderate to high PKA energies since they trigger extremely high \textit{electronic} temperatures, as locally observed on irradiated surfaces~\cite{LIN20076295}.
However, MD lacks quantum-mechanical effects required to capture electronic and electron-phonon ($\mathrm{e-ph}$) coupling effects. 
Even though MD is a classical technique, is it possible to partially incorporate electronic effects within semi-empirical approaches such as the \emph{so-called} two-temperature model ~\cite{duffy2006including,Tamm:2016,Tamm:2018,Tamm:2019,Eva:2019, Zarkadoula2014,Zarkadoula2015,ZARKADOULA2017106,Ullah_2019, Ullah:2020, Grossi:2020}. 
However, these models require several temperature-dependent material parameters that are challenging to compute and, in many cases, remain unavailable. 
This lack of parameters is more accentuated in novel materials where properties are unavailable to the scientific community. 

{\color{black}
First-principles simulations provide valuable insights into the electronic properties of materials, especially in novel compositions. 
Typically, such simulations are heavily constrained by system size and spatial complexity~\cite{Ponga:2016, Ponga:2020}.
In particular, it is challenging to accurately represent statistical representation of configurational disorderliness as in HEA cases as the number of simulation cells increases quickly with number of atoms and species. 
Indeed, if we let $N_e$ be the number of atomic species (or elements), $N_s$ be the number of sites in a multi-principal element metallic alloy, and $N_i$ be the number of atoms of the $i-$th element, and using the combination notation, the number of possible representations ($N_\mathrm{cells} $) becomes $N_\mathrm{cells} =   \displaystyle\prod_{i=1}^{N_e} \begin{pmatrix} R_i \\ N_i \end{pmatrix}$\footnote{Using the combination notation defined as $\begin{pmatrix}
N_s \\ N_1 
\end{pmatrix} =\frac{N_s!}{N_1! (N_s-N_1)!}$.}.  $R_i$ is the remaining number of sites defined as $R_i=N_s$ when i=1; else $R_i=N_s- \displaystyle \sum_{j=1}^{i-1} N_{j-1}$.
Noteworthy, the number of possible representations increases quickly with the number of sites and elements, and therefore, several realizations using large supercells are needed to obtain statistically meaningful results. 
The most computationally intensive way of tacking this issue is to use a large number of supercells, albeit there is not robust way to \emph{a priori} ensure that the number and size of supercells are sufficient to achieve accurate statistical representation. 
A more refined approach is so called special quasi-random structures (SQS)~\cite{PhysRevLett.65.353,PhysRevB.42.9622}, which has been used for studying properties of HEAs~\cite{Gao2016}. 
Despite its more systematic conceptualization, SQS still requires to appropriate choice of supercell size and clustering size, which increases with the number of principal elements.
The coherent-potential approximation (CPA)~\cite{PhysRev.156.809, PhysRev.156.1017}, in particular in its most-common practical implementations such as within the Korringa-Kohn-Rostoker (KKR) methods~\cite{KORRINGA1947392, PhysRev.94.1111}, or the  exact muffin-tin orbitals (EMTO)~\cite{PhysRevB.64.014107}, has been also frequently used approach to study HEAs~\cite{Tian2016}. 
Recently, it has been applied in an automated high-throughput framework to  study a large compositional space of HEAs~\cite{PhysRevMaterials.6.023802}. 
While CPA-based methods are highly promising to approximate a large number of materials properties, it is still a challenging task to obtain the large number of electronic and phonon  properties required to consistently obtain the temperature-dependent electron-phonon ($\mathrm{e-ph}$) coupling factor ($G_\mathrm{e-ph}$), electronic specific heat ($C_\mathrm{e}$) and electronic thermal conductivity ($\kappa_\mathrm{e}$) for a given compositional space. 
For instance, it is quite computationally heavy to obtain phonon dispersion within CPA-based approaches for a large compositional space~\cite{PhysRevB.104.024205}. 
Thus, Samolyuk et al. have used a linear mixing scheme of the constant e-ph mass enhancement parameter ($\lambda$) of Ni, Cr, and Fe for that of their equi-molar binary alloys~\cite{Samolyuk_2016}. 
A less robust, yet extremely expedient approach is the virtual-crystal approximation (VCA)~\citep{PhysRevB.61.7877}, which is conventionally applied a linear mixing scheme to atomic pseudo-potentials~\citep{HEINE19701}. 
It offer a disordered mean-field limit approximation for configurational disorderliness as in HEAs.   
By construction, VCA is missing any short-ranged ordering, or details of local chemical environment. 
However, it is highly suitable when approximating \emph{average} materials properties of disordered system such as $G_\mathrm{e-ph}$, $C_\mathrm{e}$ and $\kappa_\mathrm{e}$, which are commonly approximated by averaging over microscopic quantities. 
Despite its crude oversimplification, it has been shown that VCA is relative accurate when calculating  elastic constant, simple-phase transformation, and thermodynamic functions of HEAs~\citep{10.3389/fmats.2017.00036}.
In particular, it can provide suitable tools when trend analysis throughout a composition space due to equally present systematic errors within it. 
The VCA approximation is used in this work to compute average electronic properties of HEAs including several elements in a wide range of compositions.

This work presents a first-principles investigation of electronic properties in MPE alloys at several molar compositions. 
The temperature-dependent $G_\mathrm{e-ph}$, $C_\mathrm{e}$ and $\kappa_\mathrm{e}$ are calculated within fully first-principles approaches.
There has been previous works~\cite{Samolyuk_2016,PhysRevB.92.144309,ZARKADOULA20166,ZARKADOULA2017124}  on elemental metals and binary alloys, partially carrying the electronic temperature effects. 
In this work, the underlying quantities such as $\lambda$, the Eliashberg function, the Fermi velocity, the plasmon lifetime and the electronic thermal conductivity are directly calculated for each electronic temperature point between $100-30000$~K for each composition.
Additionally, the influence of the electronic effects in an equimolar CoCrFeNi HEA is investigated under radiation environments within MD simulations incorporating the electronic effects using the \textit{local} two-temperature molecular dynamics ($\ell$2T-MD) method~\cite{Ponga_2018,Ullah_2019}.
For the first time, our work provides accurate first-principles electronic data for MPE alloys and shows the importance of these effects in radiation damage. 
Remarkably, we have found that neglecting these effects could lead to large overestimations of the FP population (up to $\sim 20-25 \%$) even at low PKA energies of $50$~keV.
An expedient model is developed to predict vacancy formation for the compositional space of Cr-Fe-Co-Ni.
Temperature-dependent first-principles electronic, and phonon properties database, and MD defect-evolution database are made  available in the repository at Ref.~\citenum{PongaHEADatabase}.

\section{Theoretical and computational methodology}

\subsection{Calculating first-principles electronic, and phonon properties}

Let us now describe the relevant parameters of interest needed to quantify the electronic effects in multi-PE alloys. 
The temperature-dependent $\mathrm{e-ph}$ coupling factor $G_\mathrm{e-ph}$ is given by~\cite{PhysRevLett.59.1460,doi:10.1063/1.4935843}
\begin{align}\label{eq:eq1}
G_\mathrm{e-ph}=-\frac{\pi \hbar k_\mathrm{B} \lambda \left\langle \omega^2 \right\rangle}{ g(E_\mathrm{F})}\int_{-\infty}^\infty  d\epsilon \; \frac{\partial f(\epsilon,\mu,T_I^\mathrm{e})}{\partial \epsilon} g^2(\epsilon),
\end{align}
 where $k_\mathrm{B}$ is the Boltzmann constant, $g(E)$, $E_\mathrm{F}$, and $\mu$ are the electronic density of states (EDOS), the Fermi energy, and the chemical potential, respectively; $f(\epsilon,\mu,T_\mathrm{e})$ is the Fermi-Dirac distribution function. The product of $\lambda$ and the mean square phonon
frequency, $\lambda \left\langle \omega^2 \right\rangle$, is given by
\begin{align}\label{eq:eq2}
\lambda \left\langle \omega^2 \right\rangle=2 \int_0^\infty d\omega\;  \omega \alpha^2 F(\omega), 
\end{align}
 with  $\alpha^2 F(\omega)$ is the Eliashberg spectral function. 
The temperature-dependent electronic specific heat, $C_\mathrm{e}$, is given by~\cite{doi:10.1063/1.4935843}
\begin{align}\label{eq:eq3}
C_\mathrm{e}=\int_{-\infty}^\infty  d\epsilon \; (\epsilon-E_\mathrm{F}) \frac{\partial f(\epsilon,\mu,T^\mathrm{e})}{\partial T^\mathrm{e}} g(\epsilon). 
\end{align}
The final component to describe the electronic properties of materials related to heat conduction, is the temperature-dependent thermal conductivity, $\kappa_\mathrm{e}$, given by~\cite{LIN20076295},
\begin{align}\label{eq:eq4}
\kappa_\mathrm{e}= v_\mathrm{F}^2\; \tau_\mathrm{p}\; C_\mathrm{e},
\end{align}
where $v_\mathrm{F}$, and $\tau_\mathrm{p}$ are the Fermi velocity, and the plasmon lifetime, respectively, which are also implicitly dependent on the electronic temperature. 

The three central quantities in Eqs.~\eqref{eq:eq1},~\eqref{eq:eq3},~and~\eqref{eq:eq4} require accurate estimation of $g(E)$, $E_F$, $\lambda$ and $\alpha^2 F(\omega)$, $v_\mathrm{F}$ and $\tau_p$. 
Kohn-Sham density-functional theory (KS-DFT)~\cite{PhysRev.136.B864,PhysRev.140.A1133,PhysRevB.21.5469,PhysRevB.33.8822,PhysRevB.46.6671} is almost-ubiquitous first-principles approach to calculate electronic structures of solids. 
Its perturbative extension, called the density-functional perturbation theory (DFPT)~\cite{PhysRevB.55.10337,PhysRevB.55.10355}, also provides an expedient approach for obtaining zero-temperature phonon properties. 
Despite rapid advancements in underlying algorithms, and computational resources, the KS-DFT-based implementations are severely limited in terms of system size and spatial complexity, in particular for metallic systems \cite{Ponga:2016,Ponga:2020}. 
It is particularly challenging in the cases of HEAs, since it requires a large number of supercell simulations to achieve an accurate statistical representation configurational disorderliness. 
Virtual-crystal approximation (VCA) is an expedient approach to study multi-principles elements solids at the disordered mean-field limit~\cite{doi:10.1002/andp.19314010507}. 
Within VCA, the atomic pseudo-potentials~\citep{HEINE19701} of constituent elements are linearly mixed to a single pseudopotential, representing a virtual atom~\cite{PhysRevB.61.7877}.  

{\color{black}
In principle, DFT is an \emph{absolute-zero-temperature} theory. 
However, it is worth to mention that there has been considerable work on extending it to capture thermal equilibrium at finite temperatures~\cite{10.1007/978-3-319-04912-0_2}.
In the context of this work, the extreme electronic temperature are highly localized as it is generated by high-energy particles. 
Thus, the finite electronic temperature effects can be expediently included within a simple conceptual way through the smearing parameter around the Fermi level, representing $T_\mathrm{e}$, in the case of metallic systems.  
}
With that in mind, $g(E)$ is expected to have a negligible $T^\mathrm{e}$-dependence as it only affects occupancies of the electronic states at around the Fermi level as the practical DFT assign occupancies after optimizing the KS states.
It is more suitable to include the possible effects of $T^\mathrm{e}$ through $E_F$ as it does not require a large number of electronic band-structure simulations. 
The $T^\mathrm{e}$-dependent $E_F$ can be easily obtained using the zero-temperature electronic structure with  little additional computational cost. 
We refer the reader to Refs.~\citenum{Hofmann_2009}~and~\citenum{PONCE2016116} for details on how to obtain $\lambda$ and $\alpha^2 F(\omega)$, using DFPT. 
The $T^\mathrm{e}$-dependence can be partially introduced by setting the smearing parameter used for the double-delta integral during the e-ph coupling simulations.

Consistently, the averaged Fermi-velocity square, $\langle v^2_\mathrm{F}\rangle$, can approximately given by~\cite{doi:10.1080/14786435808237011}
\begin{align}\label{eq:eq5}
\langle v^2_\mathrm{F}\rangle=
\frac{\Big(\sum_m\int_{S_\mathrm{F_m}}\der{k} \left|\frac{\partial E_{m,\mathbf{k}}}{\partial \mathbf{k}}\right|^2
 \Big)}
{\Big( \sum_n \int_{S_\mathrm{F_n}}\der{k'} \Big)},  
\end{align}
assuming that the electronic bands have parabolic dispersions normal the Fermi surface.
$S_\mathrm{F_m}$, and  $E_{m,\mathbf{k}}$ are the partial Fermi surface, and the KS energy of the $m-\textrm{th}$  metallic band, respectively.  
In practice, $\langle v^2_\mathrm{F}\rangle$ is calculated on a slab with a thickness of $\Delta E_\mathrm{Fermi}$ using the in-house post-processing tool~\cite{Orhan_2019}, available in Ref.~\citenum{OrhanQEDrude}. 
It requires a well-converged electronic band structure on an extremely dense Brillouin-zone sampling.
The $T^\mathrm{e}$-dependence can be consistently included by setting $d_\mathrm{Fermi}=k_\mathrm{B}T^\mathrm{e}$~\cite{doi:10.1021/acs.jpcc.1c06110}. 
We point out that \eq{eq5} does not include the $\mathrm{e-ph}$ coupling effects, which can be introduced as the electron mass renormalization such as $v_\mathrm{F} \rightarrow v_\mathrm{F}/(1+\lambda)$~\cite{Grimvall_1976}. 

Ignoring any impurity, the plasmon lifetime of a metallic solid can be expressed as the sum of the electron-electron ($\mathrm{e-e}$) and the electron-phonon ($\mathrm{e-ph}$) scatterings terms within Matthiessen's rule such as
\begin{align}\label{eq:eq6}
\frac{1}{\tau_\mathrm{p}}=\frac{1}{\tau_\mathrm{e-e}}+\frac{1}{\tau_\mathrm{e-ph}}.
\end{align}
The $\mathrm{e-e}$ term (in the Hz unit) can be estimated using  the Fermi liquid theory at the absolute-zero temperature by~\cite{doi:10.1080/00018739300101514,PhysRevB.55.10869,DalForno2018}
\begin{align}\label{eq:eq7}
\frac{1}{\tau_\mathrm{e-e}}=\frac{m e^4 (E-E_\mathrm{F})^2}{64 \pi^3 \hbar^3 \epsilon_0 E_s^{3/2}E_\mathrm{F}^{1/2}}
\left(\frac{2\sqrt{E_sE_\mathrm{F}}}{4E_\mathrm{F}+E_s}+\arctan\sqrt{\frac{4E_\mathrm{F}}{E_s}}\right),
\end{align}
where $m$, $e$, $\hbar$, and $\epsilon_0$ are the electronic mass, the unit charge, the reduced Plank's constant and the vacuum permittivity, respectively. 
The kinetic energy due to the Thomas-Fermi screening length $q_s=e \sqrt{g(E_\mathrm{F})/\epsilon_0}$ given by $E_s=\hbar^2 q_s^2/(2m)$.
By setting $(E-E_\mathrm{F})=k_\mathrm{B}T^\mathrm{e}$, the $T^\mathrm{e}$-dependence can be consistently introduced alongside the $T^\mathrm{e}$-dependent $E_\mathrm{F}$. 
The second term in \eq{eq6} is approximately given by~\cite{PhysRevB.71.174302,Hofmann_2009}
\begin{align}\label{eq:eq8}
\frac{1}{\tau_\mathrm{e-ph}}=\frac{2\pi k_\mathrm{B} \lambda T_\mathrm{e} }{3}.
\end{align}

\subsubsection{Computational details of first-principles simulations}
First-principles simulations were performed using the Quantum ESPRESSO (QE) software~\cite{0953-8984-21-39-395502,0953-8984-29-46-465901} with the external thermodynamics  \texttt{thermo\_pw}~package~\cite{ThermoPW}. 
Fermi velocities were calculated using an in-house code available in Ref.~\cite{OrhanQEDrude}. 
The SG15 optimized norm-conserving Vanderbilt (ONCV) scalar-relativistic pseudo-potentials~\cite{PhysRevB.88.085117,SCHLIPF201536} (version 1.2) using  the Perdew-Burke-Ernzerhof (PBE) exchange-correlation functional~\cite{PhysRevLett.43.1494,Kerker_1980,PhysRevB.40.2980} were used for the four base elements (Cr, Fe, Co and Ni), as well as the additional elements (Al, Mn and Cu). 
Atomic pseudo-potential were used to obtain the virtual atoms representing MPEs alloys within VCA.
The spin-polarization has been ignored to be consistent as the spin-polarization has not been yet implemented within the quasi-harmonic approximation~\citep{Palumbo_2017} in the \texttt{thermo\_pw}~package. 
However, the relevant electronic and phonon properties are needed to approximate macroscopic quantities by averaging over the the first Brillouin zone.
Thus, while the local magnetic ordering may play a crucial role in phonon dispersions, it is expected to become less significant after averaging, in particular for HEAs~\cite{PhysRevB.87.075144,PhysRevB.96.014437,PhysRevMaterials.3.124410}.

The initial crystallographic information for the thermodynamically most-stable phases of the base elements was extracted from Ref.~\cite{doi:10.1107/S0365110X65000361}. 
Cr and Fe have body-centered cubic (BCC) structures, while Co and Ni have hexagonal closed-packed (HCP) and face-centered cubic (FCC) structures.  
The FCC Ni structure was used as a template for the initial crystal structures for the HEAs.  
Variable-cell relaxation was performed using a common high kinetic energy cutoff of $150$~Ry on a $12\times 12 \times 12$ Monkhorst-Pack-equivalent Brillouin zone sampling (MP-grid)~\cite{PhysRevB.13.5188}, and the convergence for the total energies, the total forces, and the self-consistency were set to $10^{-7}$~Ry, $10^{-6}$~Ry$
\cdot$a$_0^{-3}$ and $10^{-10}$, respectively.

To practically mimic the extreme electronic temperatures of  $\sim 30000$~K on irradiated surfaces, the temperature grid was divided into two parts such as a low temperature grid of $100-1000$~K with a step size of $100$~K, and a high temperature grid of $3000-30000$~K with a step size of $3000$~K.
The electronic temperatures were consistently introduced a smearing parameters around the Fermi levels in all simulations.
The first set of simulations were performed to obtain anharmonic thermodynamic properties were calculated for the low-$T^\mathrm{e}$ using the quasi-harmonic approximation~\citep{Palumbo_2017} within the \texttt{thermo\_pw}~package on a shifted  $12\times12\times12$ MP-grid for electronic structure simulations, and a $4\times4\times4$ MP-grid for phonon simulations. 
The $\mathrm{e-ph}$ mass enhancement parameters were calculated by following the procedure in  Refs.~\citenum{Hofmann_2009}~and~\citenum{PONCE2016116} with a dense $36\times36\times36$ MP-grid for interpolation, and a converged $12\times12\times12$ MP-grid for electronic structure simulations,  $4\times4\times4$ MP-grid for phonon simulations. 
Finally, the Fermi velocities were calculated using the in-house code starting from the band structures, calculated on a $36\times36\times36$ MP-grid. 
Temperature-dependent electronic, and phonon properties upto $3\times10^4$~K have been made publicly available in the repository at Ref.~\citenum{PongaHEADatabase}.

\subsection{$\ell2$T-MD methodology} 

The $\ell$2T-MD method, developed  by Ullah and Ponga~\cite{Ullah_2019}, provides computationally feasible and seamless approach to simulate the electronic effects in MD simulations within the two-temperature model. 
Considering a system with $N_a$ atoms, each atom is provided with an electronic temperature variable. Let  ($T_i^\mathrm{e}$)  denote the electronic temperature associated with the $i-$th atom in the system which can fluctuate locally depending on the local environments during a cascade even \footnote{We point out that the subindexes $i$ and $j$ are used in the $\ell$2T-MD method to denote different atomic sites and should not be confused with traditional notation within the DFT formulations for different orbitals.}.
At the same time, let us introduce a maximum electronic temperature $T_\mathrm{max}^\mathrm{e} $ representing an \emph{arbitrary} upper bound to the electronic temperature of the system. 
On the basis of this maximum temperature, we can map the temperature field using the transformation $\theta_i= \frac{T_i^e}{T_\mathrm{max}^\mathrm{e}}$ which maps $\theta_i \in [0,1]$ interval. 
According to Ullah and Ponga~\cite{Ullah_2019}, the $\ell$2T-MD model electronic temperature evolution using the following master equation.
\begin{align}\label{eq:eq9}
&\frac{\partial T_i^\mathrm{e}}{\partial t} = \underbrace{\frac{G_\mathrm{e-ph} }{C_e}T_\mathrm{max}^\mathrm{e} \left(\theta_i^\mathrm{lat}-\theta_i^\mathrm{e}\right)}_{\mathrm{e-ph}} \nonumber \\ 
& + T_\mathrm{max}^\mathrm{e}  \sum_{j\ne i}^{N_\mathrm{A}} 
\underbrace{
K_{ij}
\left[\theta_j^\mathrm{e}(1-\theta_i^\mathrm{e})\;e^{\theta_i^\mathrm{e}-\theta_j^\mathrm{e}} \right. \left. -\theta_i^\mathrm{e}(1-\theta_j^\mathrm{e})\;e^{\theta_J^\mathrm{e}-\theta_i^\mathrm{e}}\right]}_{\mathrm{e-e}}  \nonumber \\
\end{align}
In \eq{eq9} the time-evolution of $T_i^\mathrm{e}$ for each atom is influenced by the energy exchange energy between electrons surrounding the $i-$th atom (and denoted by ($\mathrm{e-e}$)), and a coupling between electron and phonons ($\mathrm{e-ph}$). 
$K_{ij}$ is a material property that quantifies the rate of electronic energy exchange between electrons near two adjacent atomic sites. Interestingly,  $K_{ij}$ can be linked to well-defined properties obtained either from \emph{ab-initio} simulations or experimental data and is going to be described below.

The term $G_\mathrm{e-ph}$ measures the $\mathrm{e-ph}$ scattering between electrons close to the Fermi level and phonons and determines how effectively the energy exchange between $\mathrm{e-ph}$ occurs. 
On the other hand, the term $C_\mathrm{e}$ determines how much excess energy electrons can absorb.
By definition, both of these terms are always positive for metals.
A higher $G_\mathrm{e-ph}$ leads to more rapid energy exchange between $\mathrm{e-ph}$, while a higher $C_\mathrm{e}$ allows to store more excess energy in the free electrons. 
When a sudden amount of energy is introduced in the system, there is a rapid rise in the lattice temperature ($T_i^\mathrm{lat}$), which electrons around the Fermi level can partially absorb until the electronic and lattice temperatures reach equilibrium.  

We notice that $\theta_i^\mathrm{lat} = T_i^\mathrm{lat}/T_\mathrm{max}^\mathrm{e} $ in \eq{eq9} denotes a mapped local lattice temperature that is is classically defined by
\begin{align}\label{eq:eq10}
T_i^\mathrm{lat}=\frac{2}{3 k_\mathrm{B}N_c}\sum_{\substack{j=1, \\j\neq i}}^{N_c}\frac{1}{2}m_j v_j^2,
\end{align}
$N_c$ is the number of atoms in a small cluster surrounding the $i-$th atom. The small cluster is defined within a cut-off radius, $r_\mathrm{c}$, which is equal to the  interatomic potential's cut-off used in MD~\cite{Ullah_2019}. 
$m_j$, and $v_j$ are the atomic mass and the velocity vector of the $j-$th atom. 

The pair-wise exchange-rate thermal coefficient $K_{ij}$ for the electronic temperature can be determined by a long-wave analysis to well defined thermodynamics properties \cite{Ponga_2018,Ullah_2019,Mendez:2021} and is given by
\begin{align}\label{eq:eq6}
K_{ij}=\frac{2  d  \kappa_\mathrm{e}}{C_\mathrm{e} Z b^2} = \frac{2 d  v_\mathrm{F}^2 \tau_\mathrm{p} }{Z b^2 },
\end{align}
where $d=3$ is the system dimension, $Z$ is the coordination number of the lattice without distortion, and $b$ is the Burgers vector.

\subsubsection{Computational details of classical molecular dynamics simulations}
MD and $\ell$2T-MD simulations were performed using the Large-scale Atomic/Molecular Massively Parallel Simulator (LAMMPS) software~\cite{plimpton1993fast}, into which the $\ell$2T-MD have been implemented by Ullah and Ponga~\cite{Ullah_2019}. 
The $\ell$2T-MD routine was modified to allow $G_\mathrm{e-ph}$, $C_\mathrm{e}$, and $\kappa_\mathrm{e}$ to be updated with $T_i^\mathrm{e}$. 
A binary search algorithm on a hook-up table was used to find the correct parameters range corresponding to $T_i^\mathrm{e}$, while linear interpolation was used to interpolate the values corresponding to the exact $T_i^\mathrm{e}$. 
The embedded atom model (EAM) potential by Farkas and Caro~\cite{farkas2018model} was used as the interatomic potential as it can model HEA of Co, Cr, Fe and Ni and was used previously in literature for modeling radiation damage for HEA ~\cite{DeLuigi:2021, qian2021atomistic}.  
The electronic stopping was incorporated in the model through electron stopping fix in LAMMPS~\cite{stewart2018,PhysRevB.102.024107}.
The stopping-power values were calculated using the Stopping and Range of Ions in Matter (SRIM) software~\cite{ziegler2010srim} with a minimum cut-off energy of $10$~eV.
For the sake of completeness and robustness, the minimum cut-off energies of $1$, $5$ and $10$~eV were tested for CrFeCoNi within the $\ell$2T-MD. 
The short-range interactions were modified using a ZBL potential~\cite{10.1007/978-3-642-68779-2_5} to prevent atoms from getting too close during the displacement cascades. 
The ZBL potential was smoothly linked to the EAM pair potential for distances between 0.05 nm and $0.18$~nm. This procedure does not affect the equilibrium properties.

An initial cubic simulation cell with a dimension of $L=34$~nm was used to fill using a generic crystal with a lattice parameter of $a=0.363$~nm with a total number of atoms of $3,538,944$. This cell size ensured that the radiation-induced damage was contained within the simulation cells' boundaries.
Periodic boundary conditions were employed in all three directions for all simulation cells.
The initial simulation cells of CoCrFeNi were obtained by randomly placing Co, Cr, Fe and Ni on the atomic sites.
We performed the following test to validate the random generation to ensure the correct statistical sampling. First, 512 random generations were performed in smaller samples containing $\sim500$ atoms. These samples were relaxed using molecular statics and subjecting them to zero pressure. Then, the energy per atom ($\sim-4.1232$ eV/atom) and its standard deviation ($\sim4$ meV/atom) were measured. These samples indicated a distribution close to a Gaussian of energies after minimization, as shown in \ref{fig:MDHistogram}. Subsequently, nine large cells involving $3,538,944$ atoms were relaxed under the same conditions, and their average energy ($\sim-4.1229$ eV/atom) and standard deviation were also obtained. The difference between average energies was less than 1 meV/atom, indicating a large number of chemical environments are simulated in the large sample. The energy levels correspond to the average values of the 512 different realizations suggesting proper sampling of the chemical environments. 

Then the simulation cells were subjected to energy minimization at $T=0$~K using the Polak-Ribi\`ere conjugate gradient algorithm~\cite{10.1007/BFb0099521}. 
Subsequently, an initial temperature of $T=300$ K was given to the atoms, and the system was relaxed for $30$~ps using the Nos\'e$-$Hoover thermostat to relax the hydrostatic pressure to zero. The same energy minimization and relaxation procedures were used for the Ni simulation cells.

Cascade events were initiated by randomly choosing a primary knock-on-atom (PKA) located at the center of the simulation cells. 
The PKA was given a random velocity vector corresponding to recoil energy of $50$ keV in an NVE ensemble. 
For the $\ell$2T-MD simulation, the average lattice and the average electronic temperatures ($T_\mathrm{lat}$, and $T_\mathrm{e}$, respectively) were initially set at $T=300$~K before energizing the PKA. 
A variable time step was used with a maximum time step of $\leq 0.01$~fs for the cascade simulations to limit the maximum distance for every time-step moved by the PKA atom to $0.005$~\AA.
At each MD step, between $10$ to $15$ $T_\mathrm{e}$ integration time-steps of \eq{eq9} were performed. 
The total time of the cascade simulations was about $32$~ps, which was sufficient for the FPs to reach steady states. 
Each cascade simulation was repeated $\sim 20$ times with different random PKA directions and HEA structures to ensure sufficient statistical representation.
To ensure this, the mean values of the total numbers of defected atoms ($N_\mathrm{def}$) and Frenkel pairs ($N_\mathrm{FP}$) was analyzed for the randomly selected subsets of the cascade simulations with respect to the number of simulations, shown in \fig{Ndefmax-MDSimNum-Conv} and \fig{NFPmax-MDSimNum-Conv}, respectively.
Our analysis indicates that the total number of cascade simulations for each case reaches to convergence within $\pm 5\%$ accuracy of the overall mean value of all available simulations for each case.

The common-neighbor analysis (CNA)~\cite{TSUZUKI2007518} was used to identify formations of defect clusters. 
The Ovito software was used for visualization of the defect structures~\cite{Stukowski:2010}, and the DXA algorithm was used for dislocation loops analysis \cite{stukowski2012automated}. 
Finally, the Wigner-Seitz defect analysis was performed to determine the vacancies and the self-interstitial atoms (SIA) using the Ovito software. 

The related database of defect evolutions in Ni, and CrFeCoNi within the conventional MD, and the $\ell2$T-MD has  been made publicly available in the repository at Ref.~\citenum{PongaHEADatabase}.

\section{Results and Discussion}
We start with  thermodynamic and mechanical stability assessments of our base HEA (CrFeCoNi) and its PEs.
The mixing Gibbs free energy ($G_\mathrm{mix}$) is an ubiquitous indicator for thermodynamic stability~\cite{e21010068}. 
In \fig{Xx4-GFE}, the mixing Gibbs free energy ($G_\mathrm{mix}$) and the individual contributions to it are shown up to $1000$~K.
The configurational-entropy contribution is the predominant term, leading to thermal stability at a low temperature. At the same time, the mixing electronic and the mixing vibrational Helmholtz free energies ($F_\mathrm{el}^\mathrm{mix}$, and $F_\mathrm{vib}^\mathrm{mix}$, respectively) are negligible (see the Supplementary Materials (\textbf{SM}) for the details to calculate the free energies, and the configurational entropy).  
Addition to the thermodynamic stability, the elastic and dynamic stability is crucial to assess the mechanical stability. 
The elastic stability is assessed using the Born-Huang-stability criteria~\cite{Born:224197,PhysRevB.90.224104} (see the \textbf{SM} for the necessary and sufficient condition for cubic symmetry).
The dynamic stability is assessed by  analyzing phonon dispersions~\cite{PhysRevB.97.134114}. 
CrFeCoNi is both elastically stable according to the Born-Huang-stability criteria and dynamically stable as it has neither negative nor soft phonon modes~\cite{PhysRevB.97.134114}. 
\begin{figure}[H]
\centering
\includegraphics[width=0.45\textwidth]{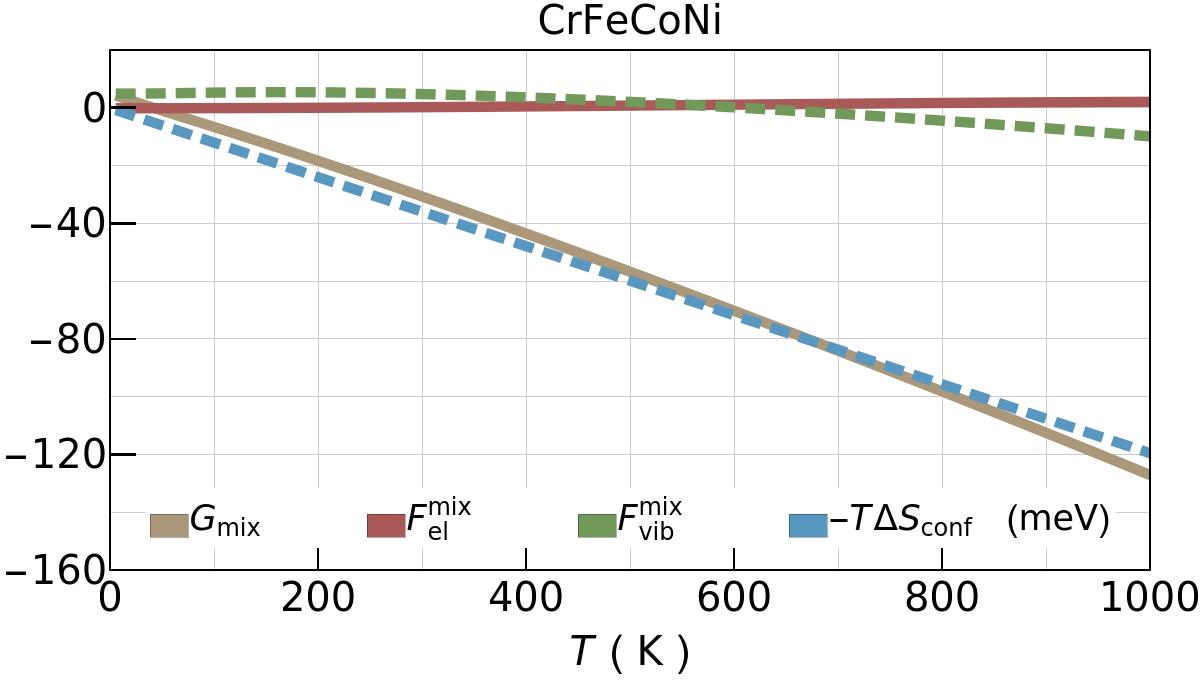}
\caption{
Temperate dependent mixing Gibbs free energy ($G_\mathrm{mix}$) for CrFeCoNi in RSS. Mixing electronic, and vibrational Helmholtz free energies ($F_\mathrm{el}^\mathrm{mix}$ and $F_\mathrm{vib}^\mathrm{mix}$, respectively), and the configurational-entropy contribution ($-T \Delta S_\mathrm{conf}$) also shown in the plot. See \textbf{SM} for calculations details.
}
\label{fig:Xx4-GFE}
\end{figure}

\begin{figure*}[t]
\centering
\subfloat{\includegraphics[width=0.9\textwidth]{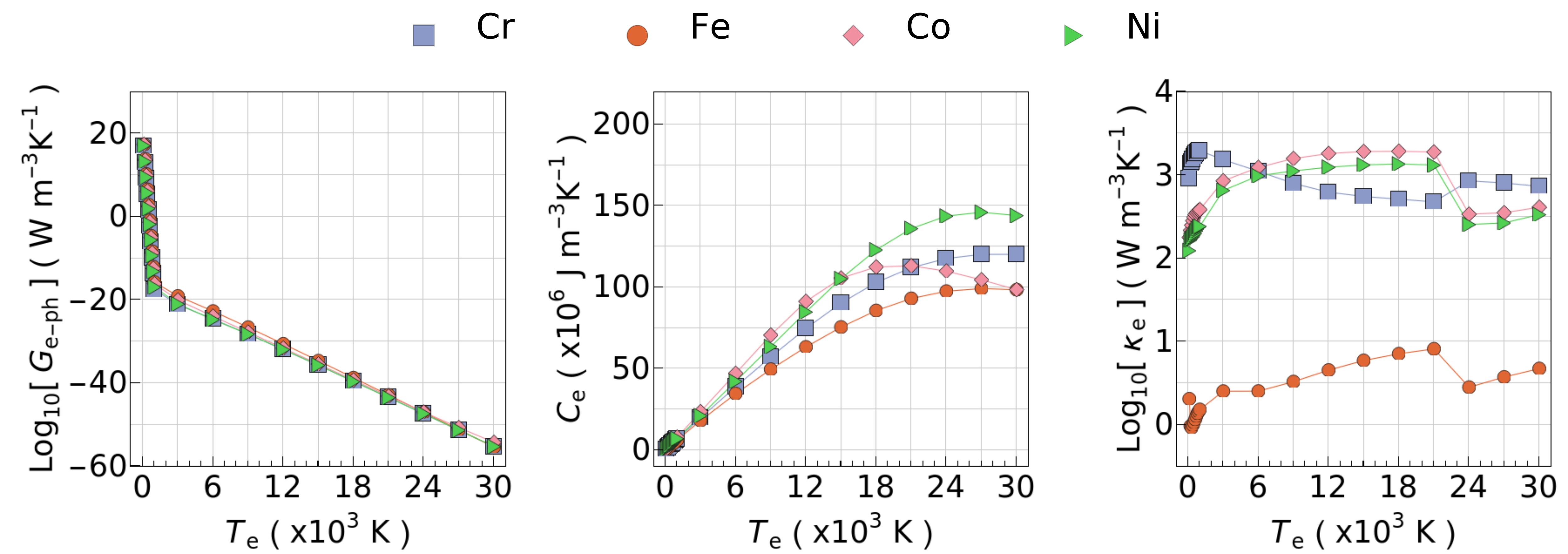}}

\subfloat{\includegraphics[width=0.9\textwidth]{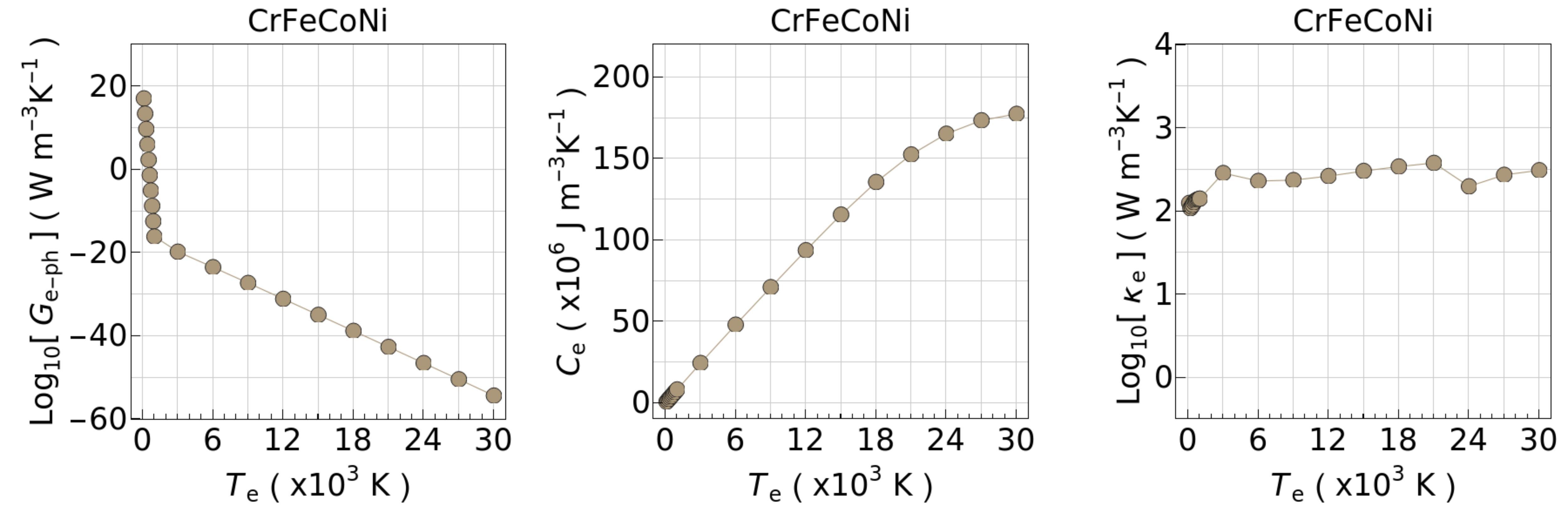}}
\caption{Temperature dependence of the $\mathrm{e-ph}$ coupling factor ($G_\mathrm{e-ph}$), the electronic specific heat ($C_\mathrm{e}$), and the electronic thermal conductivity ($\kappa_\mathrm{e}$) of the base PEs, and the FCC CrFeCoNi.
}
\label{fig:PE-Xx4-EPh-Prop}
\end{figure*}
In \fig{PE-Xx4-EPh-Prop}, $G_\mathrm{e-ph}$, $C_\mathrm{e}$, and $\kappa_\mathrm{e}$ are depicted for the four pure elements, and their equimolar FCC RSS. 
$G_\mathrm{e-ph}$ exhibits quite similar trends for CrFeCoNi and the pure elements, exponentially vanishing with the increasing temperature. 
The integral term in \eq{eq1} predominantly sets this trend when the individual terms are further analyzed as shown in Figs.~\ref{fig:PE-EPh-Prop-Comp}~and~~\ref{fig:Xx4-EPh-Prop-Comp}. 
However, the non-integral terms for each system roughly average to the same order and determine magnitudes of $G_\mathrm{e-ph}$. 
It becomes in order of $1$ at around $500$~K, indicating that electrons become too fast for $\mathrm{e-ph}$ scattering.  
On the other hand, $C_\mathrm{e}$ almost-linearly increases with the increasing $T^\mathrm{e}$.
This results to a rapid $G_\mathrm{e-ph}/C_\mathrm{e} \rightarrow 0$ for $T^\mathrm{e}$ above $500$~K. 
This indicates that there will be very little energy transfer from electrons to lattice when $T_i^\mathrm{e} >> T_j^\mathrm{lat}$.
On the other hand, despite also decreasing $\kappa_\mathrm{e}/C_\mathrm{e}$ ratio, there will still be energy dissipation, mediated solely by electrons.
This scenario is desirable to keep FPs formation by dissipating excess energy among electrons while avoiding further energizing the lattice. 

Comparing $C_\mathrm{e}$ of CrFeCoNi with the pure elements, we observed that CrFeCoNi has a higher $C_\mathrm{e}$, hinting that electrons have a better capacity to absorb energy compared to the pure metals. 
This effect could be important in radiation environments as the plasmons take on more thermal energy, which is removed from the lattice and thus could result in less FP formation. 
At lower temperatures, Cr exhibits higher $\kappa_\mathrm{e}$ compared to other PEs and CrFeCoNi, while CrFeCoNi, Co, and Ni performs better at higher electronic temperatures (i.e., $T^\mathrm{e} \gtrsim 6000$~K). 
The trends in $\kappa_\mathrm{e}$ are predominantly determined by $v_\mathrm{F}$, portrayed in Figs.~\ref{fig:PE-EPh-Prop-Comp}~and~~\ref{fig:Xx4-EPh-Prop-Comp}, as well as $\lambda$ as it renormalize $v_\mathrm{F}$ by including $\mathrm{e-ph}$ coupling effects. 
This effect leads to remarkable poor performance for Fe as it has the slowest $v_\mathrm{F}$ with the largest $\lambda$.
Despite its not-so-superior $\kappa_\mathrm{e}$ when compared to those of the PEs, CrFeCoNi has a much higher $C_\mathrm{e}$. 
It leads to a much smaller $K_{ij} \propto = v_\mathrm{F}^2\tau_\mathrm{p}$, indicating that while its electrons could hold more energy for the same electronic temperature, they are less effective in dissipating it compared to the PEs. 
In summary, CrFeCoNi is expected to absorb excess lattice energy more effectively via $\mathrm{e-ph}$ coupling compared to the PEs, while less effective to dissipate it via $\mathrm{e-e}$ coupling except for Fe.

\subsection{Compositional space of non-equimolar Cr-Fe-Co-Ni random solid solutions}
Higher $G_\mathrm{e-ph}$ and $C_\mathrm{e}$ are the key factors for a superior radiation-damage resistance in HEA.   
Thus, a potential strategy to improve radiation damage properties is to tune molar fractions of PEs to further improve these properties.
We investigate the compositional space of the Cr-Fe-Co-Ni RSS.
For a systematic yet feasible investigation, we vary each PE at a time on a molar-fraction grid of $\mathrm{x}= [0, 0.05, 0.10, 0.15]$ while keeping the remaining three elements equimolar among themselves.
The $\mathrm{x}=0$ cases are the equimolar 3-PEs cases; however, the non-CrFeCoNi cases are called the non-equimolar cases for simplicity. 
By varying each PE one by one, the individual effect of each PE can be partially assessed, although reducing a single PE automatically increases the molar fraction of the remaining three.     

The first step is to assess the relative stability of the non-equimolar RSS. 
In \fig{Yy-Xx3-DGFE}, the Gibbs free energy differences ($\Delta G$) of the RSS compared to the CoCrFeNi are illustrated up to $1000$~K. 
Reducing Cr or Fe leads to thermally less stable compositions by increasing $G$ while reducing Ni has an opposite effect.
Reducing the molar-fraction of Co has a less significant effect on the thermal stability, and this is consistent with previous studies carried out in five PE alloys \cite{10.3389/fmats.2021.816610}.
At room temperature ($300$~K), a reduction up to $20\%$ of Cr or Fe, or a reduction of Co only leads to $\Delta G$ in the order of $1$~meV, which is in the order of the accuracy of the practical DFT simulations. 
Reduction of Ni already leads to better thermal stability; thus, it is thermally more favorable in any case. 
The non-equimolar RSS are also elastically stable according to the Born-Huang-stability criteria. 
They are also dynamically stable except for CoCrFe (Ni$_\mathrm{x}$(CoCrFe)$_{1-\mathrm{x}}$ for $\mathrm{x}=0$) as it has negative phonon modes at around the X high-symmetry point in the first Brillouin zone of the primitive FCC unit cell. 

\begin{figure*}[t]
\centering
\includegraphics[width=1.0\textwidth]{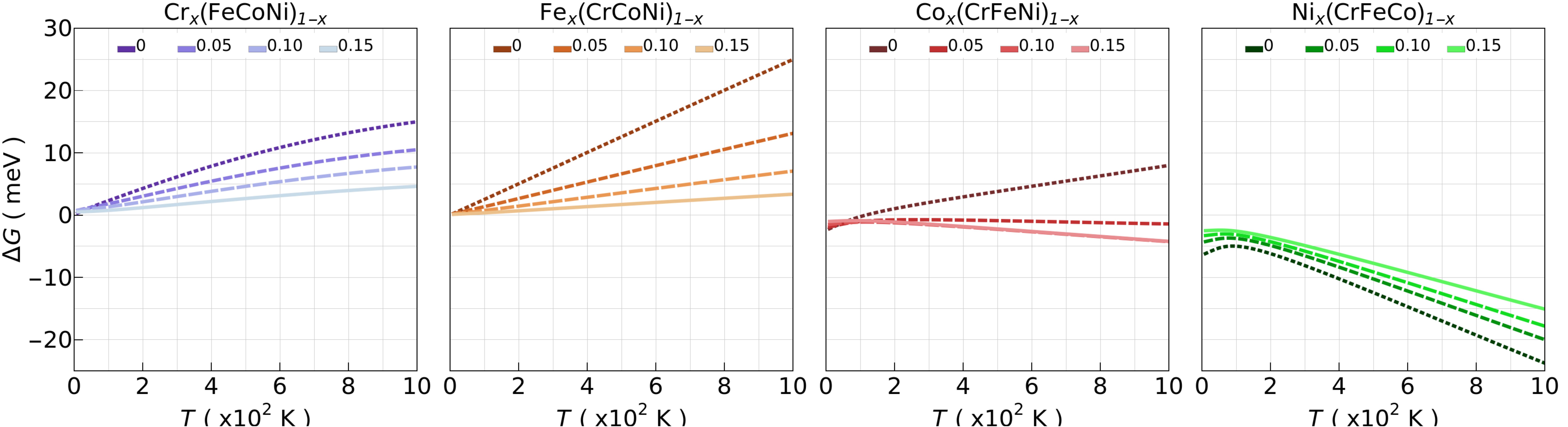}

\caption{Temperature dependent Gibbs free energy differences ($\Delta G$) for non-equimolar compositions compared to the CrFeCoNi random solid solution.
}
\label{fig:Yy-Xx3-DGFE}
\end{figure*}

In \fig{Yy-Xx3-EPh-Prop}, dimensionless ratios $\overline{G}_\mathrm{e-ph}^\mathrm{x}$, $\overline{C}^\mathrm{x}_\mathrm{e}$ and $\overline{\kappa}^\mathrm{x}_\mathrm{e}$ for non-equimolar compositions are shown to assess their relative performance in dissipating excess energy. 
The dimensionless ratios are computed as the ratio of $G_\mathrm{e-ph}$, $C_\mathrm{e}$ and $\kappa_\mathrm{e}$ of the non-equimolar RSS to those of CoCrFeNi . 
Reducing Cr concentration (first column in \fig{Yy-Xx3-EPh-Prop}) quickly worsen $G_\mathrm{e-ph}$ and $C_\mathrm{e}$ starting at around the room temperature. 
It indicates that the lower Cr concentration can lead to less effective energy exchange between $\mathrm{e-ph}$, shown to play a crucial role in promoting defect recombination and reducing defect clustering. 
Reducing Fe or Co concentrations (central columns in \fig{Yy-Xx3-EPh-Prop}) have subtle effects on $G_\mathrm{e-ph}$, $C_\mathrm{e}$ and $\kappa_\mathrm{e}$.
On the other hand, a $25-50 \%$ reduction of Cr significantly improves purely electronic heat dissipation, hinted by much higher $\kappa_\mathrm{e}$. 
However, the comparative analysis carried out later on in Section \ref{Section:MD} for Ni and CrFeCoNi indicates that the $\mathrm{e-e}$ term in \eq{eq9} has a lesser effect on the defect evolution. 
At the lower temperatures ($T<1000$~K) (the inset figures in \fig{Yy-Xx3-EPh-Prop}), lower Ni concentrations improve $G_\mathrm{e-ph}$ with a less significant loss in $C_\mathrm{e}$.
Unlike Cr, lower Ni concentrations drastically worsen $\kappa_\mathrm{e}$ ($\sim 50\%$) at lower temperatures. 
The composition-dependent relative changes in $G_\mathrm{e-ph}$ are determined mainly by ratios of the electron-phonon mass enhancement parameters ($\lambda$), while the ratios of the Fermi velocities primarily determine the general trends in $\kappa_\mathrm{e}$ ($v_\mathrm{F}$), shown in Figs.~\ref{fig:Cr-Xx3-EPh-Prop-Comp}~-~\ref{fig:Ni-Xx3-EPh-Prop-Comp}.

\begin{figure*}[t]
\centering
\includegraphics[width=1.0\textwidth]{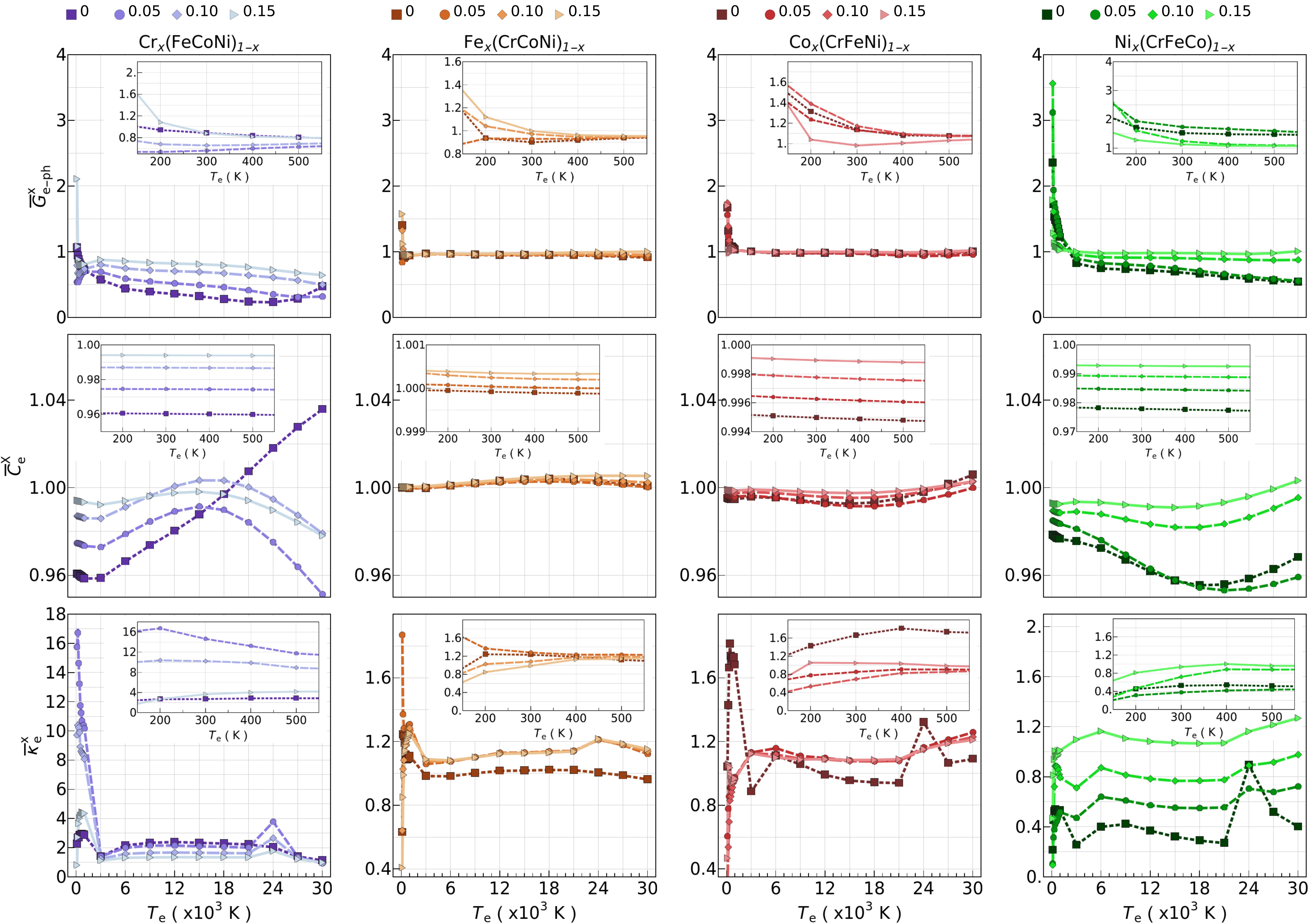}

\caption{Temperature dependence the electronic property ratios for non-equimolar compositions compared to CrFeCoNi. 
The $\mathrm{e-ph}$ coupling factor ratio ($\overline{G}_\mathrm{e-ph}^\mathrm{x}$), the fractional electronic specific heat ($\overline{C}_\mathrm{e}^\mathrm{x}$), and the  fractional electronic thermal conductivity ($\overline{\kappa}_\mathrm{e}^\mathrm{x}$) fractions for non-equimolar random-solid solutions compared to CrFeCoNi.
}
\label{fig:Yy-Xx3-EPh-Prop}
\end{figure*}

\subsection{Equimolar additions of Al, Mn, or Cu}
Another strategy to tailor the electronic and phonon parameters is to add other elements rather than exploring non-equimolar compositions.
In \fig{Xx5-DGFE}, the relative thermal stabilities of equimolar AlCrFeCoNi, CrMnFeCoNi (also known as Cantor alloy), and CrFeCoNiCu are shown with respect to CrFeCoNi.
While introducing Al or Mn leads to better thermal stability, CrFeCoNiCu becomes less thermally stable. 
However, it is still thermally stable in its own merit as indicated its mixing Gibbs free energy, shown in \fig{CrMnFeCoNi-Gmix} using the thermal data for the elemental Cu from Ref.~\citenum{doi:10.1021/acs.jpcc.1c06110}
While all three RSS are elastically stable according to the Born-Huang-stability criteria, only CrMnFeCoNi and CrFeCoNiCu are dynamically stable as AlCrFeCoNi has negative phonon modes at around L high-symmetry point in the first Brillouin zone of the primitive FCC unit cell. 

\begin{figure}[H]
\centering
\includegraphics[width=0.45\textwidth]{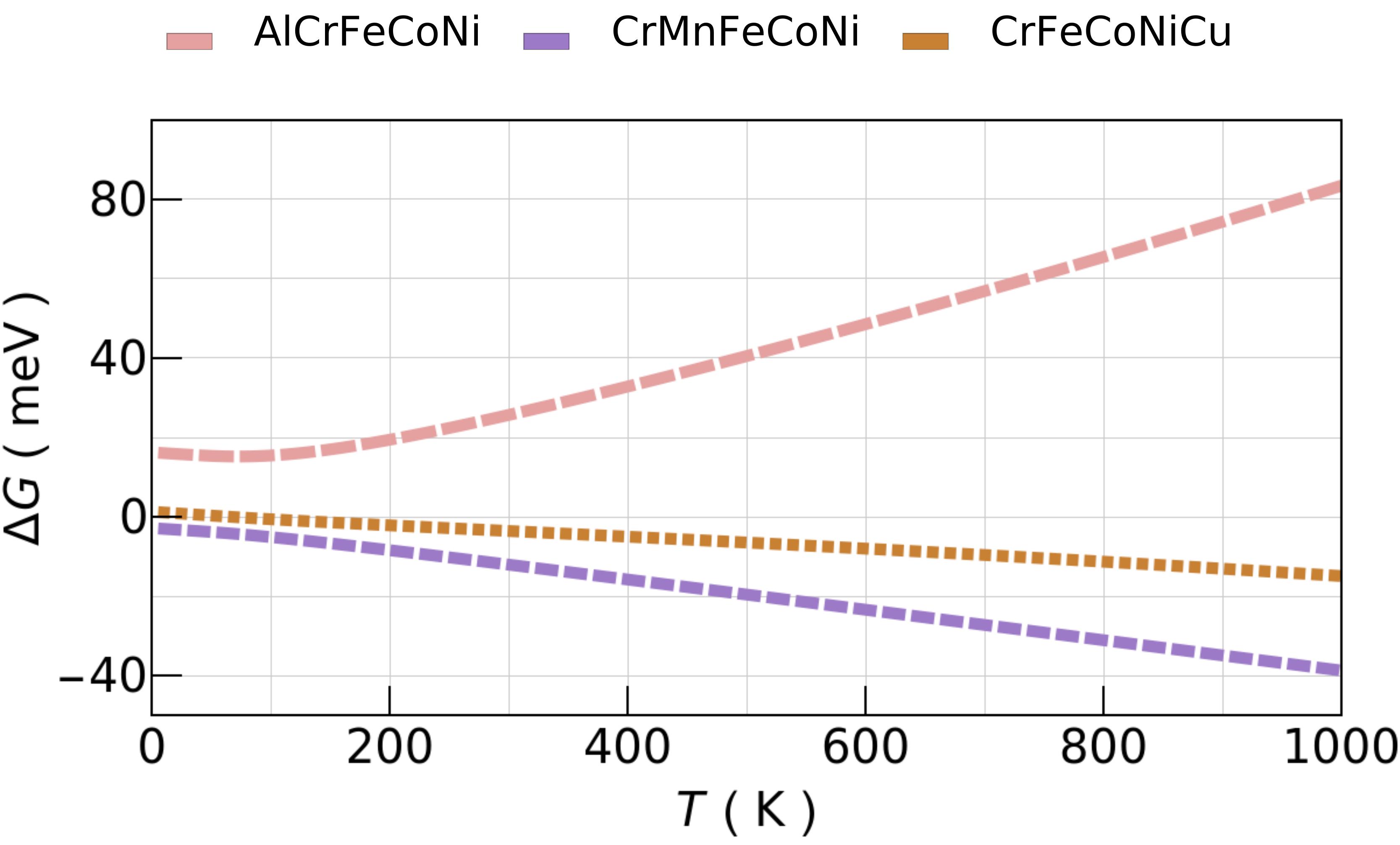}
\caption{
Temperature dependent Gibbs free energy differences ($\Delta G$) for equimolar AlCrFeCoNi, CrMnFeCoNi, and CrFeCoNiCu compared to the CrFeCoNi.
}
\label{fig:Xx5-DGFE}
\end{figure}

{\color{black}
In \fig{Xx5-EPh-Prop}, the relative performance of the electronic ($\overline{G}_\mathrm{e-ph}^\mathrm{Al/Mn/Cu}$, $\overline{C}^\mathrm{Al/Mn/Cu}_\mathrm{e}$ and $\overline{\kappa}^\mathrm{Al/Mn/Cu}_\mathrm{e}$) properties  compared to CrFeCoNi are shown.
The very first observation is that the addition of Al drastically increases $G_\mathrm{e-ph}$~\footnote{Note that the scale for Al is represented at the left of the plot and is several times higher than for the other two elements.}, as well as substantially increases $C_\mathrm{e}$ and $\kappa_\mathrm{e}$ by more or less $\sim2$ times compared to those of CrFeCoNi.
On the other hand, the addition of Mn has almost-no effect on $G_\mathrm{e-ph}$, $C_\mathrm{e}$, or $\kappa_\mathrm{e}$. 
Finally, the addition of Cu worsens $G_\mathrm{e-ph}$ and improves $\kappa_\mathrm{e}$ $\sim 1.5$ times while almost no effect on $C_\mathrm{e}$. 

The drastic increase in $G_\mathrm{e-ph}$ of AlCrFeCoNi is mostly due to the integral term in \eq{eq1}, as well as higher $\lambda \left\langle \omega^2 \right\rangle$ and lower $g(E_\mathrm{F})$, as displayed in Figs.~\ref{fig:AlCrFeCoNi-EPh-Prop-Comp}~and~\ref{fig:Xx5-EDOS-PDOS}. 
The higher $\lambda$ values can be mainly attributed to lower averaged mass as phonon modes are normalized with the average mass of the unit cell. 
On the other hand, introducing Mn, which is \textit{most-alike} to CrFeCoNi in terms of its averaged atomic mass and VEC, has the most negligible effects on the EDOS and phonon DOS (PDOS); thus, $G_\mathrm{e-ph}$ is almost identical except at the very low temperatures. 
In the case of CrFeCoNiCu, the lower $\lambda$ alongside negligible increment in other significant terms in \eq{eq1}, shown in \fig{CrFeCoNiCu-EPh-Prop-Comp}, leads to a lower $G_\mathrm{e-ph}$.

In the case of $C_\mathrm{e}$, the EDOS is the principal figure of merit, as imposed by \eq{eq3} (in particular, due to $\mu \approx E_\mathrm{F}$)
When comparing the EDOS of the four systems in the top-right panel of ~\fig{Xx5-EDOS-PDOS}, the EDOS of CrFeCoNi, CrMnFeCoNi, and CrFeCoNiCu are pretty similar. 
On the other hand, AlCrFeCoNi has lower yet more spread-out EDOS, leading to higher $C_\mathrm{e}$. 
With that in mind, the higher $\kappa_\mathrm{e}$ of AlCrFeCoNi is due to its higher $C_\mathrm{e}$, as well as its faster free electrons despite shorter-living, shown in \fig{AlCrFeCoNi-EPh-Prop-Comp}.
Without any contribution from $C_\mathrm{e}$, the almost-identical $\kappa_\mathrm{e}$ of CrMnFeCoNi is due to counter-balancing of changes in $v_\mathrm{F}$ and $\tau_\mathrm{p}$.
On the other hand, CrFeCoNiCu has faster as well as longer living plasmons, leading to slightly higher $\kappa_\mathrm{e}$. 
As $\tau_\mathrm{p}$ is inversely proportional to $\lambda$ via \eq{eq8}, which is the predominant term in \eq{eq6}, a higher $\lambda$ leads to a lower $\tau_\mathrm{p}$.
Similarly, $v_\mathrm{F}$ is renormalized by $(1+\lambda$), a higher $\lambda$ lower it. 
However, a steeper band dispersion can compensate for this reduction, as in the case of AlCrFeCoNi. 

When introducing additive elements, the relative changes in $G_\mathrm{e-ph}$ and $\kappa_\mathrm{e}$ are determined by both electronic and phonon properties, while $C_\mathrm{e}$ is purely electronic.
In the case of $\kappa_\mathrm{e}$, it is also dependent on fine details of electronic band dispersion and Fermi surface.
}

\begin{figure*}[t]
\centering
\includegraphics[width=0.9\textwidth]{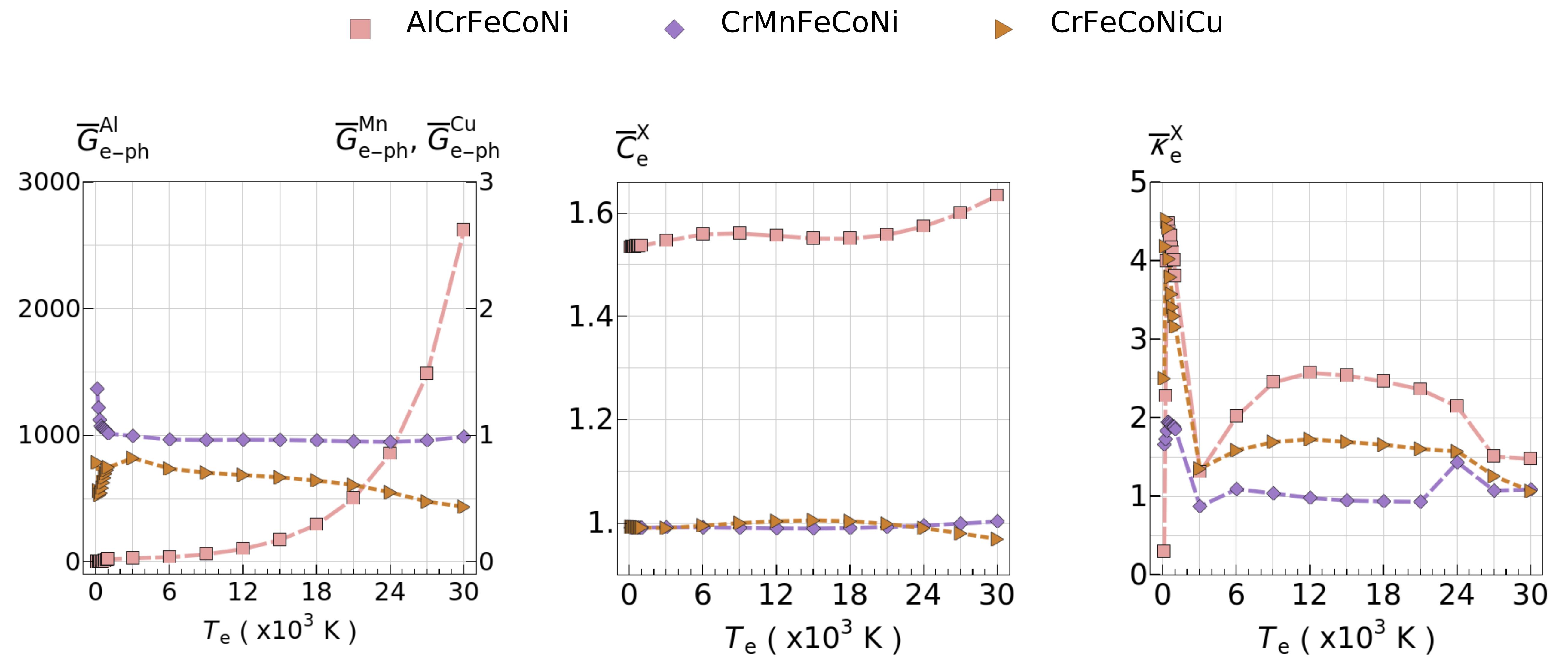}
\caption{Temperature dependence of the electron-phonon coupling factor ($G_\mathrm{e-ph}$), electronic specific heat ($C_\mathrm{e}$), and electronic thermal conductivity ($\kappa_\mathrm{e}$) fractions of  AlCrFeCoNi, CrMnFeCoNi, and CrFeCoNiCu compared to CrFeCoNi. Note that $G_\mathrm{e-ph}^\mathrm{Al}$ scale is shown at the left of the plot.
}
\label{fig:Xx5-EPh-Prop}
\end{figure*}

\subsection{Benchmark: Defect evolution in Ni and CrFeCoNi} \label{Section:MD}
Despite its relatively well-scalability, accurate MD simulations are still computationally challenging, particularly to achieve a reasonable statistical representation of HEAs.
Thus, the elemental Ni and CrFeCoNi are chosen as the benchmark cases to simulate the defect evolution.  

\begin{figure}[H]
\centering
\includegraphics[width=0.45\textwidth]{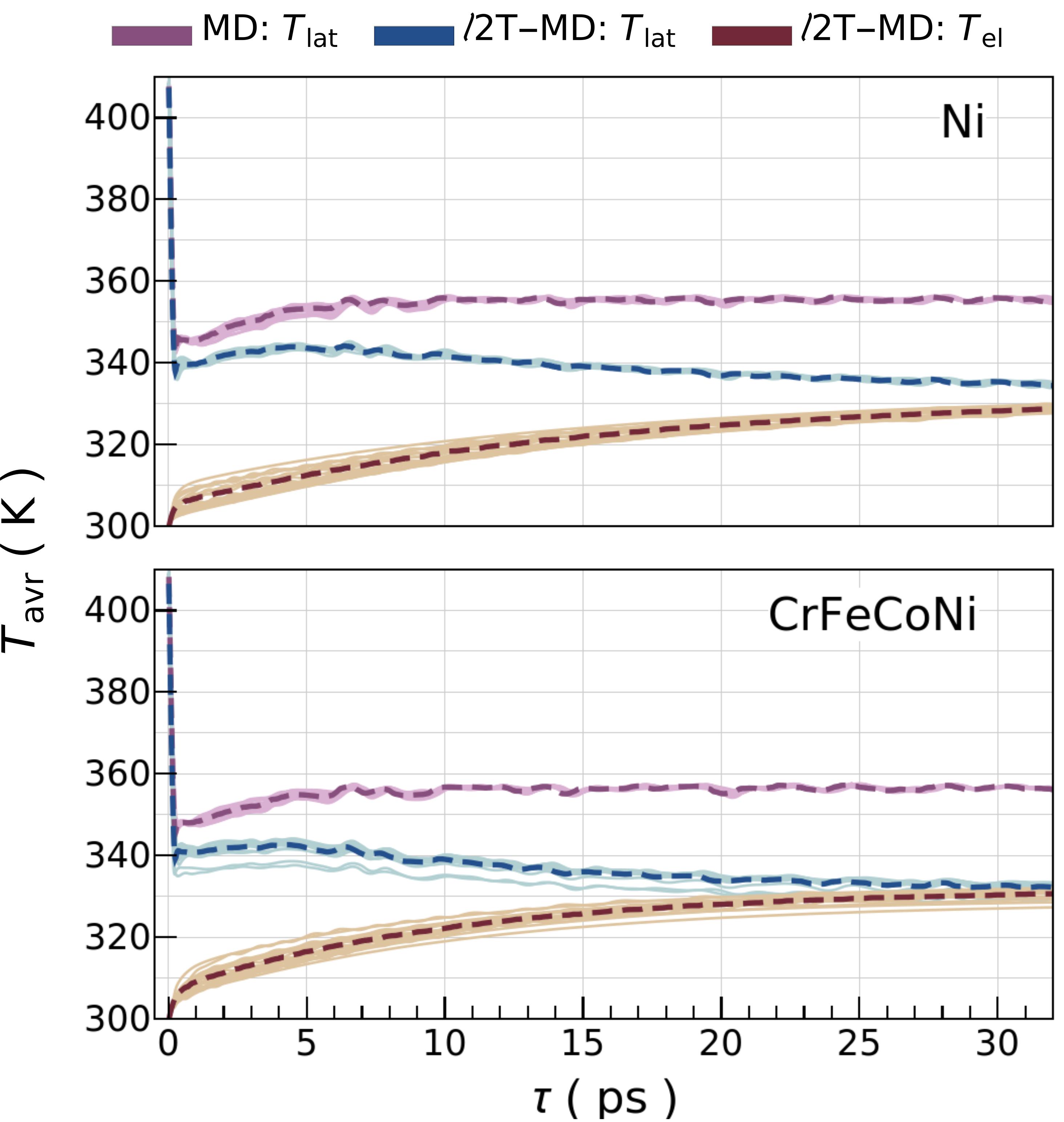}
\caption{
Time evolution of the average lattice ($T_\mathrm{lat}$) and electronic ($T_\mathrm{e}$) temperatures for Ni and CrFeCoNi within the conventional MD and $\ell$2T-MD for a 50 keV PKA.
The mean values of the repeated eight simulations are shown with dashed lines. 
}
\label{fig:Tlat-and-Te-Evolution}
\end{figure}

We start by analyzing the time evolution of the average temperatures depicted in \fig{Tlat-and-Te-Evolution}. For both materials, the initial average $T_\mathrm{lat}$ spike to $\sim 408$~K from the initial $300$~K due to the additional kinetic energy of PKA. 
We also notice that for the $\ell2$T-MD simulations, the average electronic temperature also shown in \fig{Tlat-and-Te-Evolution} remains at 300 K at $t=0$ ps.
For classic MD and $\ell2$T-MD simulations, the lattice temperature showed a quick dip at around $\sim 0.25$~ps in $T_\mathrm{lat}$ (listed in ~\tab{SM-T-Curves}) as the PKA transfer energy to the surrounding atoms and rapidly generated defects. 
After that, the lattice temperature quickly stabilizes near a steady-state value that is very similar for MD simulations (355 K and 356 K for Ni and CoCrFeNi, respectively). 

On the other hand, examining the $\ell2$T-MD simulations, we observed a much different behavior. 
For those simulations incorporating the electronic effects, the lattice temperature continuously reduced its value as time went by, whereas the electronic temperature monotonically increased. 
These two temperatures eventually reached equilibrium (at 334 K and 332 K for Ni and CoCrFeNi, respectively) at around $t=32$ ps. 
We also observed that the standard deviation of these values was around $\pm2$ K, illustrating a very robust statistical response of the multiple replicas used in the study. 
The analysis of the average temperature of the system with time suggests subtle differences between Ni and CoCrFeNi, and thus, it is beneficial to investigate the evolution of the maximum temperatures during the simulations.

\begin{figure}
\centering
\includegraphics[width=0.45\textwidth]{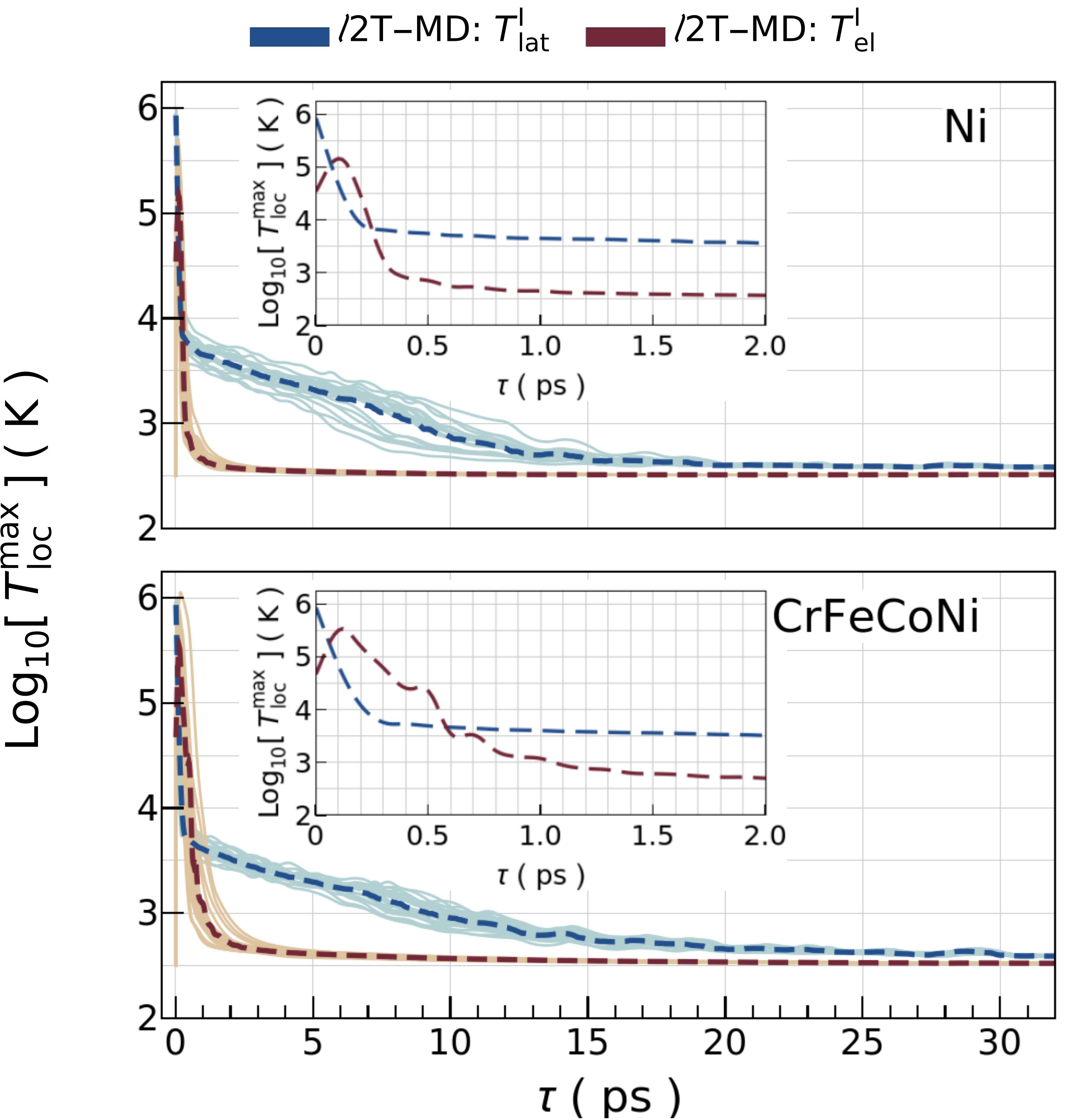}
\caption{
Time evolution of the maximum local lattice temperatures ($T_\mathrm{lat}^I$), electronic temperatures ($T_\mathrm{e}^I$) of Ni and CrFeCoNi within the $\ell$2T-MD for a 50 keV PKA.
The mean values of the repeated simulations are shown with dashed lines. 
}
\label{fig:TlatI-and-TeI-Evolution}
\end{figure}

\fig{TlatI-and-TeI-Evolution} compares the maximum local lattice and electronic temperatures during the cascade simulations within $\ell2$T-MD simulations. 
At first glance, we observed that the peak of lattice temperature was the same for both materials, whereas the peak for the electronic temperature was different (see insets in \fig{TlatI-and-TeI-Evolution}).
The time to reach the maximum electronic temperature ($\sim0.05$ ps) is comparable with plasmon lifetime $\tau_p$ and could hint at ballistic electronic heat conduction in  near the PKA for both materials. 
While ballistic heat conduction is neglected in the electronic subsystem, it is considered through the explicit integration of the equation of motion in the lattice. We point out that the $\ell2$T-MD implementation allows for second sound simulations, and a computational implementation is provided in Refs.~\cite{Ponga_2018,Mendez:2021}.
We also observed for CoCrFeNi several oscillations in the maximum electronic temperature, hinting at local plasmon excitations during the PKA trajectory. 
As a result, and considering that the width of the peak was much longer for the CoCrFeNi compared to Ni, the lattice temperature dropped faster for CoCrFeNi than Ni. 
These facts hint at a better energy exchange between $\mathrm{e-ph}$ for CoCrFeNi compared with the pure metal. Indeed, the energy exchange between electrons and phonon is more significant in the case of CrFeCoNi as indicated more prolonged survival of $T^\mathrm{e}$ before reaching steady temperatures (at around $30$~ps) compared to those of Ni, reaching steady temperatures at around $15$~ps. 

\begin{figure*}
\centering
\includegraphics[width=0.9\textwidth]{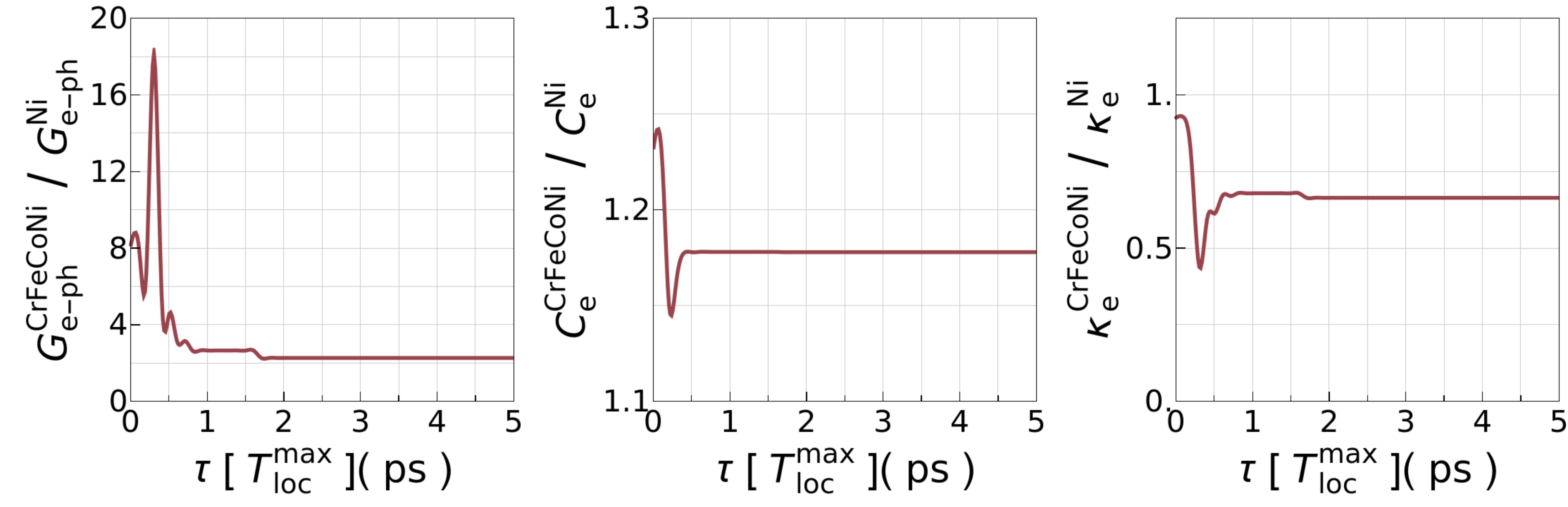}
\caption{Time evolution of the $\mathrm{e-ph}$ coupling factor ($G_\mathrm{e-ph}$), the electronic specific heat ($C_\mathrm{e}$), and the electronic thermal conductivity ($\kappa_\mathrm{e}$) ratios between Ni and CrFeCoNi.
}
\label{fig:Ni-Xx4-EPh-Prop-Comparison}
\end{figure*}

To better understand this energy exchange, \fig{Ni-Xx4-EPh-Prop-Comparison} shows a graphical comparison of the local electronic properties between CrFeCoNi and Ni for the first $t=2$ ps using the maximum electronic temperature. Remarkably, the ratio between $G_\mathrm{e-ph}$ for CrFeCoNi and Ni is much greater than one below $t=0.5$ ps, and eventually converged to two; suggesting a much faster exchange of thermal energy from the lattice to the electrons. Moreover, $C_\mathrm{e}$ is about $20\%$ greater for CrFeCoNi compared to Ni although some variations are observed  below $t=0.5$~ps. Only the thermal conductivity of the high entropy alloy is less than Ni (about $70\%$ of Ni), but it seems to have a much less critical impact on the thermodynamics behavior of the two subsystems.

\subsubsection{Electronic effects in defect formation and recombination}
Having analyzed the evolution of the electronic and lattice temperatures during the simulations, we now investigate the impact on defect formation when electronic effects are included. 
 \fig{Defected-Atom-Number} depicts the time evolution of the defected atoms and FP in the simulations. 
 Defected atoms are defined as those atoms that do not belong to the FCC lattice. 
Qualitatively, the trends are very similar for defected atoms and FPs. An initial high-peak was observed at the earlier stages of the simulations, i.e., $\tau _\mathrm{def}^\mathrm{max} = 0.31$~ps (listed in \tab{Defect-Formation}). 
A monotonic decay of defected atoms and FP followed until the number stabilized after $\tau^\mathrm{steady} =8-10$ ps. 
The number of defected atoms and FP was higher for all MD simulations compared to the $\ell2$T-MD simulations, and also for Ni compared to CoCrFeNi (cf. $N_\mathrm{def}^\mathrm{steady}$ and $N_\mathrm{FP}^\mathrm{steady}$ in \tab{Defect-Formation} columns one and two, and three and four).
At the initial stages of the cascade events, the steep decrease in $T_\mathrm{lat}$ leads to a rapid defected-atom generation as indicated by the sharp increase in the total number of defected atoms ($N_\mathrm{def}$) in \fig{Defected-Atom-Number}.
The reduced number of FP compared to defected atoms indicates that during the early stages of the PKA, some defects form FP while some are only partially defected due to lattice distortions or recombine quickly due to the high local thermal energy. 
Nevertheless, besides reducing significantly after a few ps, FPs reached steady-state for all simulations after $\sim10$ ps.

As shown in \tab{Defect-Formation} the trends are analogous as the number of defected atoms with more FP for MD compared to $\ell2$T-MD and for Ni to CoCrFeNi. 
Interestingly, comparing the FP at the end of the simulations, a difference of $\sim26\%$ was observed for Ni, and this is comparable with previous studies for pure metals \cite{Zarkadoula2014,Zarkadoula2015,ZARKADOULA2017106,Ullah_2019}. Difference can be attributed to the temperature dependence implemented in our work compared to previous ones, where usually $C_e$ and $\kappa_e$ are modeled as a linearly dependent on $T_e$. However, as shown in \fig{PE-Xx4-EPh-Prop}, this is only true for moderate temperature ranges and both quantities saturate at high temperatures.
Comparing the FP for CoCrFeNi, the incorporation of electronic effects resulted in  $105 \pm 4$ FPs, whereas classical MD showed $126 \pm 8$, or a difference of $20\%$ between the two approaches. 
The significantly higher $G_\mathrm{e-ph}$ of CrFeCoNi compared to that of Ni, shown in \fig{Ni-Xx4-EPh-Prop-Comparison}, significantly promote higher defect recombination, leading to lower FPs at steady-state. 

It is evident from \fig{Defected-Atom-Number} that the electronic effects shorten the time scale to reach the steady states FPs.
Thus, we conclude that the electronic effects play a more critical role in determining the final number of FPs after the PKA, and these effects are more critical than those made by the elastic distortions present in high-entropy alloys. 
To better explain this finding, we computed the SIA  ($s_f^1$) and vacancy formation  ($v_f^1$) energies in the HEA and compared to Ni and these results are shown in \fig{SIA-Formation-Energy-PE} and \fig{Vac-Formation-Energy-PE} and listed in \tab{Defect-Formation-SM}. 
For instance, the vacancy formation energy in CoCrFeNi has a Gaussian distribution owing to the elastic distortions. However, when analyzed on average, we observed that the formation energy $v_f^1\approx1.64$ eV which is very close to one in Ni, $v_f^1=1.56$ eV. 
These small differences suggest that the energetic cost of generating a vacancy is not strongly affected by the elastic distortions and thus, cannot drastically reduce FPs formation during the PKA event. 

These findings are in line with previous results obtained by Deluigi \emph{et al.} \cite{DeLuigi:2021} where the effect of the elastic distortions was investigated and compared to an equivalent mean-field discrete sample.  
Noteworthy, Deluigi \emph{et al.} \cite{DeLuigi:2021} concluded that the elastic distortion plays a minor role in the generation of FP during low energy PKA events. 
Even though these authors did not include electronic thermal effects, we observe similar trends here. 
These results illustrate the importance of the electronic effects in MD and justify their inclusion besides increasing the simulations' computational cost.

\begin{figure*}[t]
\centering
\includegraphics[width=0.45\textwidth]{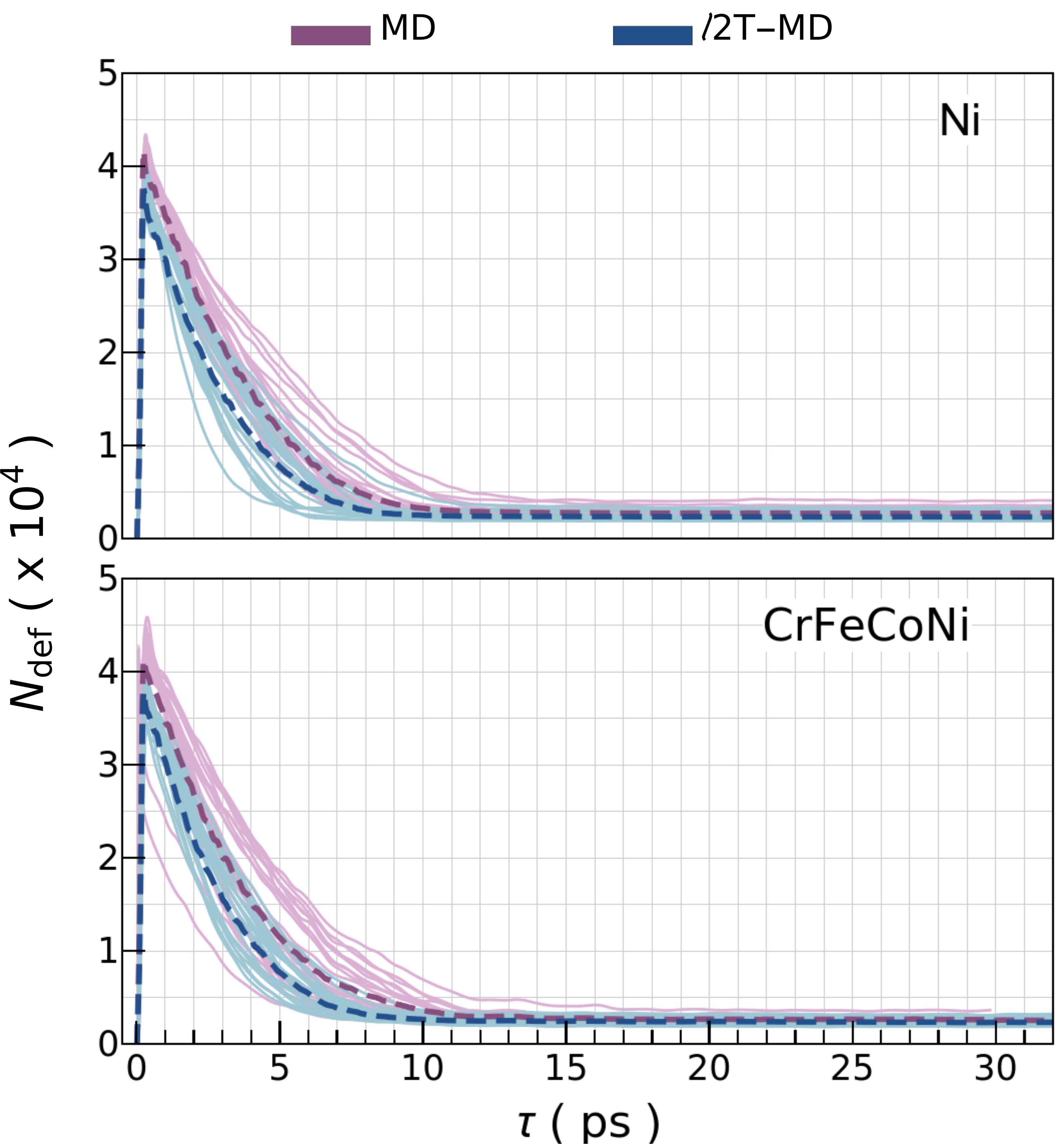}\hfill
\includegraphics[width=0.45\textwidth]{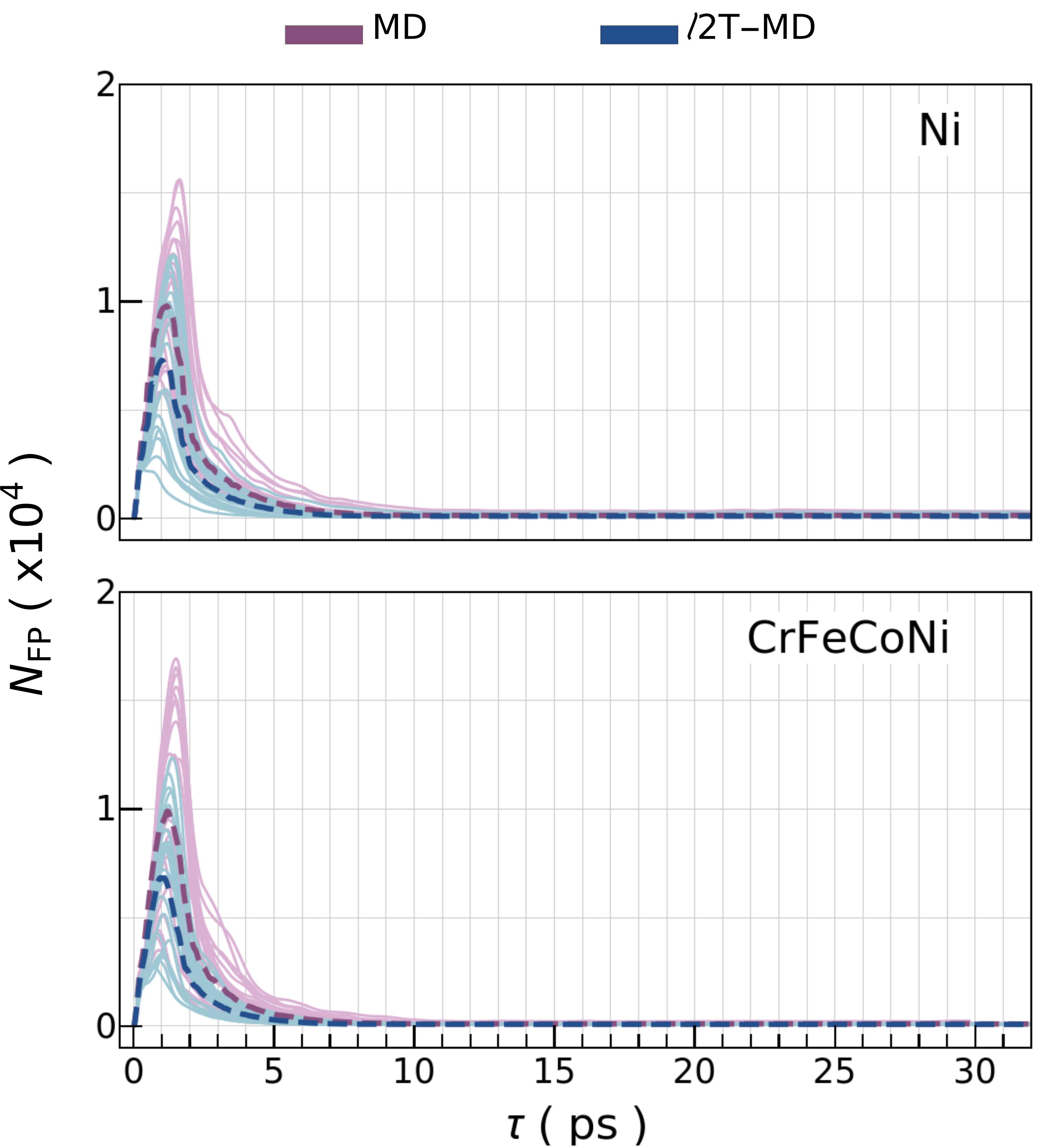}
\caption{Time evolution of the total numbers of defected atoms ($N_\mathrm{def}$) and Frenkel pairs ($N_\mathrm{FP}$) in Ni, and CrFeCoNi within the conventional MD and the $\ell$2T-MD simulations for a 50 keV PKA.  
Each simulation was repeated eight times to ensure sufficient statistical representation of cascade events.
}
\label{fig:Defected-Atom-Number}
\end{figure*}

For the sake of gaining further insight, the cascade simulations for CrFeCoNi  within the $\ell$2T-MD were repeated for the minimum cut-off energy for the electronic stopping, listed in \tab{Defect-Formation-ElStopping}.
The minimum cut-off energies ($1$, $5$, and $10$~eV) do not have any significant effects on the numbers of the defected atoms and/or Frenkel pairs.

\subsubsection{Electronic effects in defect clustering}
Another important factor in comparing the classical MD and $\ell2$T-MD simulations is the size of the defected cluster resulting after the cascade simulations. 
To this end, we graphically represented the cluster size (as denoted by the defected atoms in different defect clusters in the simulations) for twenty clusters as shown in \fig{Defect-Cluster-Size} and also visualized in \fig{FP-Clustering}. 
Again, results for MD and $\ell2$T-MD for Ni and CoCrFeNi are compared.

\begin{figure}[h]
\centering
\includegraphics[width=0.45\textwidth]{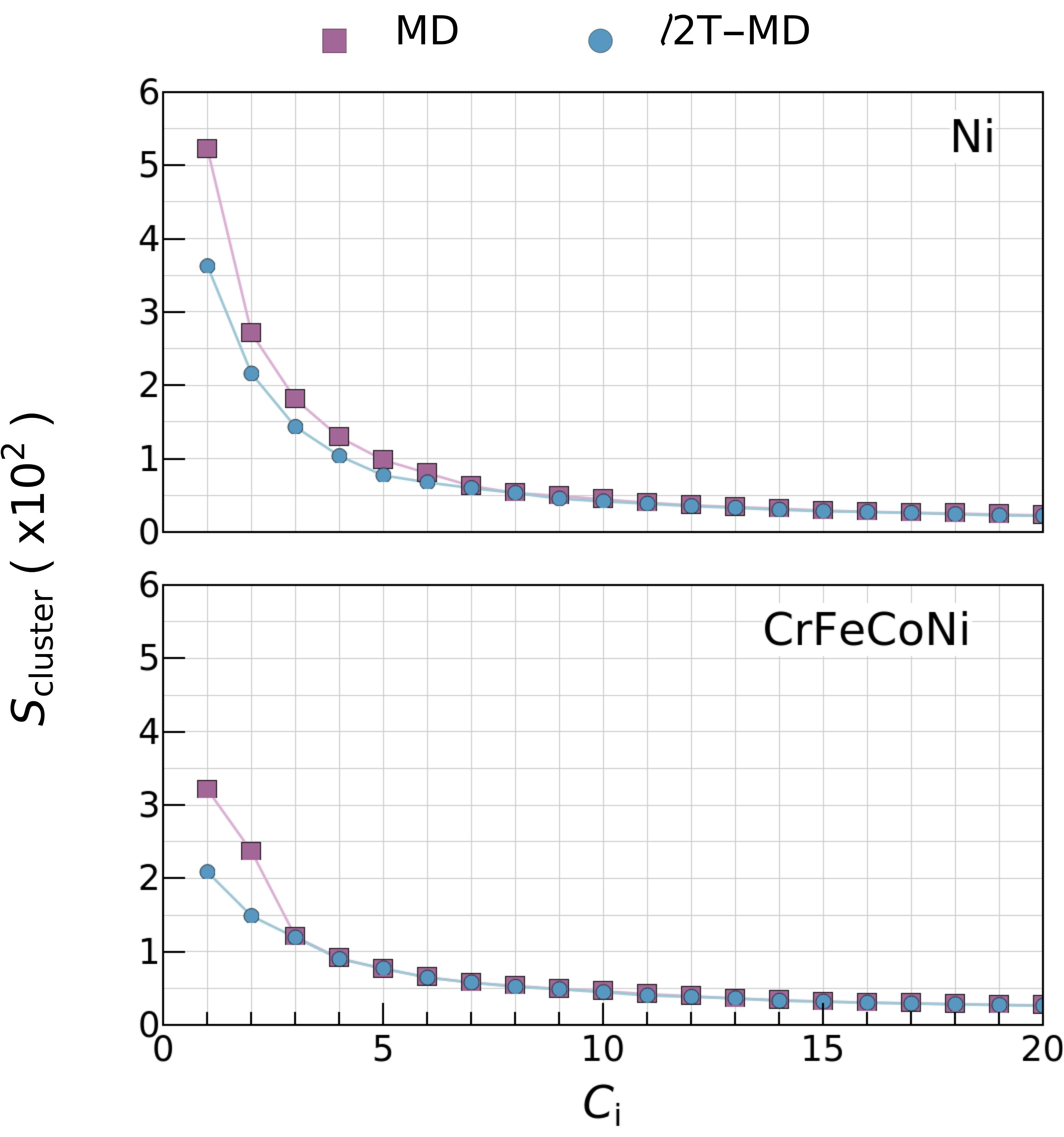}
\caption{
Comparison of the defect cluster size ($S_\mathrm{cluster}$) for each cluster in descending order of size ($C_i$) in Ni, and CrFeCoNi within the conventional MD and the $\ell$2T-MD simulations for a 50 keV PKA. 
}
\label{fig:Defect-Cluster-Size}
\end{figure}

Ni generates larger defect clusters without the electronic effects, as much as $\sim 1.5$ times larger ones than those of CrFeCoNi. 
When the electronic effects are included, this ratio rises to $\sim 2$.
Smaller $S_\mathrm{cluster}$ in CoCrFeNi compared to Ni can be attributed to better energy exchange between $\mathrm{e-ph}$ ($G_\mathrm{e-ph}$) and heat capacity ($C_\mathrm{e-ph}$). Remarkably, the lower values of thermal conductivity ($\kappa_\mathrm{e}^\mathrm{CrFeCoNi}$, displayed in \fig{Ni-Xx4-EPh-Prop-Comparison}) of the high entropy alloy compared to Ni play a much more modest role in determining the FPs and cluster size.
Despite different lattice distortion, the electronic effects reduce $S_\mathrm{cluster}^\mathrm{max}$ by $\sim 34\%$ in CrFeCoNi compared a reduction of $\sim 30\%$ in Ni. 
\begin{table*}[t]

\renewcommand{\arraystretch}{1.5} \setlength{\tabcolsep}{12pt}
\begin{center}

\begin{tabular}{llllll}

\hline \hline

      &   \multicolumn{2}{c}{\underline{Ni}}    && \multicolumn{2}{c}{\underline{CrFeCoNi}}  \\ 

      & MD &  $\ell2$T-MD && MD &  $\ell2$T-MD \\ \cline{2-3} \cline{5-6} 

$N_\mathrm{def}^\mathrm{max}$ & 40587 (402)   &  37105 (397)  & & 42066 (345)   & 37441 (412)  \\

$\tau_\mathrm{def}^\mathrm{max}$  & 0.31 (0.01)  & 0.31 (0.01)  & & 0.31 (0.02)  & 0.30 (0.01)  \\[0.25cm]

$N_\mathrm{def}^\mathrm{steady}$ &  2877 (102) & 2467 (83) & & 2754 (80) & 2483 (82) \\

$\tau_\mathrm{def}^\mathrm{steady}$  & 12.53 (0.54) & 11.70 (0.56) & & 15.13 (0.61) & 14.65 (0.74) \\[0.25cm]

$N_\mathrm{FP}^\mathrm{steady}$ &  165 (15) & 131 (11) & & 126 (8) & 105 (4) \\

$\tau_\mathrm{FP}^\mathrm{steady}$  & 9.64 (0.50) & 9.43	(1.10) &&  9.19 (0.31) & 9.72 (0.64) \\[0.25cm]
 \hline \hline

\end{tabular}

\end{center}

\caption{The mean values and standard error (in parenthesis) of the  maximum number of defected atoms at the peak of the thermal spike ($N_\mathrm{def}^\mathrm{max}$), at steady state ($N_\mathrm{def}^\mathrm{steady}$), their corresponding times (in Picoseconds units) ($\tau_\mathrm{max}$, and $\tau_\mathrm{steady}$, respectively), and the number of Frenkel pairs ($N_\mathrm{FP}^\mathrm{steady}$), obtained within the conventional MD and the $\ell$2T-MD simulations for a 50 keV PKA.}

\label{tab:Defect-Formation}
\end{table*}

\fig{FP-Fraction-PE}  exhibits the steady-state fractions of FPs per principal elements are shown for CrFeCoNi.
Even though at the initial stages of the cascade simulation, the FPs are equally distributed in all elements (not shown in the figure), the distribution quickly shifts and reaches the distribution as illustrated in \fig{FP-Fraction-PE}.
As clearly portrayed in the graphical representation, Cr disproportionately forms more FPs while Fe and Co contribute less to the total FPs population. 
This disparity in the FPs proportion is simply due to lower vacancy and SIA-formation energies of Cr compared to other PEs, as shown in \fig{SIA-Formation-Energy-PE} and \fig{Vac-Formation-Energy-PE}. 
The different  FPs fractions open up the question of exploring the compositional space of CrFeCoNi for tuning the electronic and phonon properties to improve the radiation-damage resistance.   

To better explain the distribution of FPs in a steady-state, let us use the available information and a simple mechanistic model to predict these fractions. 
This prediction is insightful because it can help guide non-equimolar HEAs' design for radiation resistance. 
First, let us define the average vacancy and SIA formation energy denoted as $\overline{v}^f = \sum_i^{N_e} \mathrm{x}_i v^f_i$ and $\overline{s}_f =\sum_i^{N_e} \mathrm{x}_i s^f_i$, respectively, with $\mathrm{x}_i$ the atomic molar fraction of the HEA. 
At the same time, let us now assume that the di-vacancy and di-SIA formation energies can be computed. 
Next, the probably that a FP formation of a given element is given by the following biased probability, $\rho_i = \exp(-\Delta v^f_i/\overline{v}^f-\Delta s^f_i/\overline{s}^f -\Delta v^{2f}_i/\overline{v}^{2f}-\Delta s^{2f}_i/\overline{s}^{2f})$. 
In the biased probability function, the arguments $\Delta$ are taking with respect to the average value, i.e., $\Delta v^f_i = \overline{v}^f - v^f_i$. 
The relative changes in formation energies shift the probabilities to generate more FPs of those elements with lower than the average values. 
Next, the fraction of FPs can be obtained by using the atomic molar fraction of the HEA considered, i.e., $\mathrm{x}^\mathrm{FP}_i = \mathrm{x}_i \rho_i$. 
Using the values provided above and the described model, we obtained for an equimolar HEA the following concentration of FP defects by elements, Co$^\mathrm{FP}=0.185$, Cr$^\mathrm{FP}=0.436$, Fe$^\mathrm{FP}=0.124$, and Ni$^\mathrm{FP}=0.255$, which is pretty approximate to the values found by MD while some discrepancy for Co and Cr is seen for $\ell2$T-MD. 
Thus, while the number of defects can be estimated well with the available models~\cite{dpa-model}, the proposed procedure can be used to estimate the FPs distribution per element.

\begin{figure}
\centering
\includegraphics[width=0.45\textwidth]{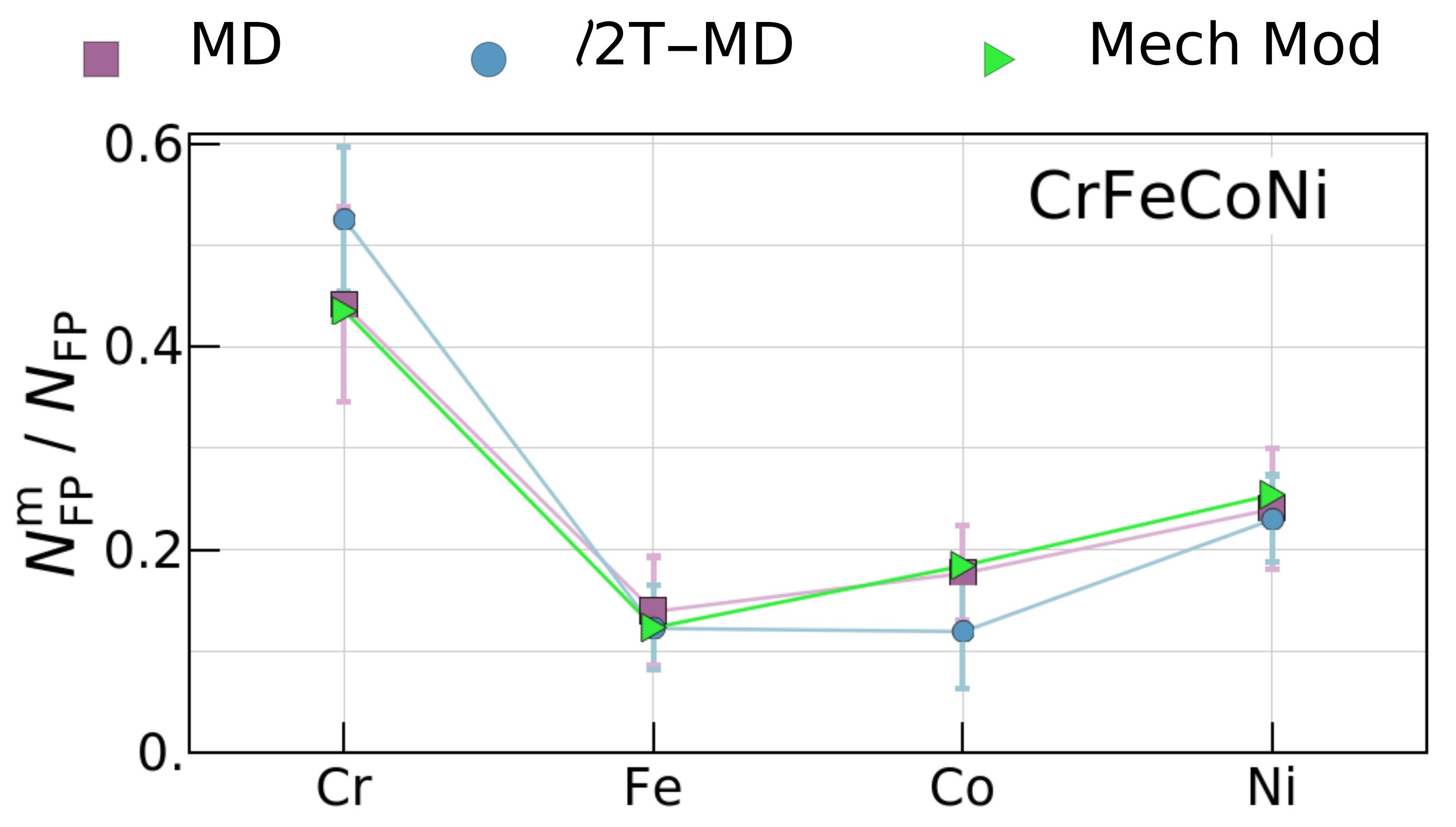}
\caption{Fractional FPs of the principal elements in CrFeCoNi after a $50$~keV PKA obtained  with MD (purple), the $\ell2$T-MD (blue). A simplistic mechanistic model results are also included (green).}
\label{fig:FP-Fraction-PE}
\end{figure}

\section{Conclusion}
This work presented a systematic study of temperature-dependent electronic and phonon properties crucial during defect formation in HEAs in radiation environments.
The FCC-phased CrFeCoNi RSS was chosen as the central HEA due to its frequent template material to develop other compositions.
It was shown that CrFeCoNi has higher $G_\mathrm{e-ph}$ and $C_\mathrm{e}$ throughout the studied electronic temperate range (up to $3\times10^4$~K) compared to its constituent elements. 
Using $\ell2$T-MD, we demonstrated that CrFeCoNi was more effective in dissipating excess energy during cascade events via local plasmons, leading to lesser defect formation and higher defect recombination than the elemental Ni.
Moreover, this work demonstrated the crucial roles of electronic and phonon effects during cascade events.

Analysis of individual affinity for defect formation of the PEs in CrFeCoNi showed that the elements such as Cr and partially Ni tend to form more FPs due to their relatively lower $v_f^1$ energies than with the remaining elements.
It motivated us to explore the non-equimolar RSS to improve the radiation-damage resistance further. 
A set of representative non-equimolar RSS compositions were investigated to understand electronic and phonon properties. 
Lower Cr or Ni concentration (keeping the remaining three PEs equimolar among themselves) lead to slightly lower $G_\mathrm{e-ph}$ and higher $\kappa_\mathrm{e}$.
Reduction of Fe or Co had negligible effects on $G_\mathrm{e-ph}$. 
Moving away from the non-equimolar case led to relatively poorer performance in $\mathrm{e-ph-}$assisted heat dissipation, while it improved energy exchange among electrons of neighboring atoms.

The addition of Al, Mn and Cu to CrFeCoNi was investigated as another strategy. 
We demonstrated that the equimolar addition of Al substantially increased $G_\mathrm{e-ph}$, which is a critical factor for plasmon excitations and to reduce defect formation.
While the equimolar addition of Mn slightly improved $G_\mathrm{e-ph}$, the equimolar addition of Cu worsens it.
On the other hand, adding any of these elements substantially improved $\kappa_\mathrm{e}$.
These general trends indicated that the addition of lighter elements compared to the averaged atomic mass improves $G_\mathrm{e-ph}$, considering that atomic mass is used when normalizing phonon modes.
On the other hand, the addition of atoms with more differing VEC led to higher $\kappa_\mathrm{e}$ due to higher $v_\mathrm{F}$, despite shorter-living plasmons when adding lighter elements.

Electronic properties were then integrated into the $\ell2$T-MD model to investigate their effect in FPs formation. We found that the electronic properties play a critical role in determining the steady-state FPs number after the PKA event. 
Neglecting these effects lead to errors of about $\sim 20-25\%$ between classical MD and $\ell2$T-MD at 50 keV PKA energy. We also investigated the fraction of FPs per principal element in the HEA. 
We found that elements with smaller vacancy formation energies (i.e., Cr and Ni) resulted in more FPs, whereas elements with higher formation energy (i.e., Fe and Co) showed fewer FPs. 
This finding suggests the possibility of exploring non-equimolar HEA compositions to obtain optimal radiation resistance combining optimal electronic and lattice properties. 

\section{Acknowledgment}
We acknowledge the support of New Frontiers in Research Fund (NFRFE-2019-01095) and from the Natural Sciences and Engineering Research Council of Canada (NSERC) through the Discovery Grant under Award Application Number 2016-06114.
 M.H. gratefully acknowledges the financial support from the Department of Mechanical Engineering at UBC through the Four Years Fellowship. 
This research partially was supported through computational resources and services provided by Advanced Research Computing at the University of British Columbia.

\end{document}


\author{Okan K. Orhan}
\author{Mohamed Hendy}
\author{Mauricio Ponga} \email[Corresponding author: ]{mponga@mech.ubc.ca}
\affiliation{Department of Mechanical Engineering, University of British Columbia, 2054 - 6250 Applied Science Lane, Vancouver, BC, V6T 1Z4, Canada}

\title{Supplementary Materials \\ ''Electronic effects on the radiation damage in high-entropy alloys''}

\maketitle

\section{Phase-stability assessment}
The mixing Gibbs free energy ($G_\mathrm{mix}$) is given for a solid with $M$ principal elements (PEs) by~\cite{ROJAS2022162309}
\begin{align}\label{eq:seq1}
G_\mathrm{mix}=H_\mathrm{mix}+F_\mathrm{mix}- TS_\mathrm{conf},
\end{align}
where $H_\mathrm{mix}$, $F_\mathrm{mix}$, and $S_\mathrm{conf}$ are the enthalpy of mixing, the mixing Helmholtz free energy, and the configurational entropy.
Using the the enthalpy of mixing of binary bulk metallic glasses ($H_{ij}$), $H_\mathrm{mix}$ is approximately given by~\cite{doi:10.1080/21663831.2014.985855},
\begin{align}\label{eq:seq2}
H_\mathrm{mix}=\sum_{i<j}4c_ic_jH_{ij},
\end{align}
where $M$, and $c_i$ are the number of PEs, and the molar fraction of the $i-$th PE. 
$H_{ij}$ can be available within the  Miedema model in Ref.~\citenum{Takeuchi2005}.
The second term in \eq{eq1} is given by~\cite{e21010068}
\begin{align}\label{eq:seq3}
F_\mathrm{mix}=F_M-\sum_i^M c_i F_i,
\end{align}
where $F_\mathrm{M}$, and $ F_i$ are the Helmholtz free energy of the $M$-PEs solid, and its $i^\mathrm{th}$ PE, respectively. 
For a non-magnetic pristine metal, the Helmholtz free energy is the sum of the electronic, and vibrational contributions, $F=F_\mathrm{el}+F_\mathrm{vib}$. 
The electronic part is given by~\cite{WANG20042665,landaystat,10.3389/fmats.2017.00036}
\begin{align}\label{eq:seq4}
F_\mathrm{el}=
\int_{-\infty}^\infty d\epsilon\; g(\epsilon)f- \int_{-\infty}^{E_\mathrm{F}} d\epsilon\; \epsilon g(\epsilon)
+k_\mathrm{B} T \int_{-\infty}^\infty d\epsilon\; g(\epsilon)\left[f \ln(f)+(1-f ) \ln(1-f) \right],
\end{align}
where $k_\mathrm{B}$, and $E_\mathrm{F}$ are the  Boltzmann constant, and the Fermi energy, respectively; $f=f(\epsilon,T_\mathrm{e})$, and $g(\epsilon)$ are the Fermi-Dirac distribution function, and the electronic density of states (DOS)
The vibrational part is given by~\cite{RevModPhys.74.11,SHANG20101040}
\begin{align}\label{eq:seq5}
F_\mathrm{vib}(T)=k_\mathrm{B}T \int_0^\infty d\omega \; \ln\left[2\; \sinh \left(\frac{\hbar \omega }{2 k_\mathrm{B}T}\right)\right]p(\omega),
\end{align}
where  $p(\omega)$ is the phonon DOS.
Finally, the configurational entropy is given within the the Stirling approximation by~\cite{Santodonato2015}
\begin{align}\label{eq:seq6}
S_\mathrm{conf}^M=-R\;\sum_i^{N} c_i \ln(c_i),
\end{align}
where $R$ is the gas constant.

The necessary and sufficient conditions within the Born-Huang-stability criteria~\cite{Born:224197}for the elastic stability of cubic systems are given by~\cite{PhysRevB.90.224104}
\begin{align}\label{eq:seq7}
\mathbb{C}_{11} - \mathbb{C}_{12} > 0,\quad \mathbb{C}_{11} + 2\; \mathbb{C}_{12} > 0 \; \mathrm{and} \quad \mathbb{C}_{44} > 0, 
\end{align}
where $\mathbb{C}_{ij}$ are the elements of the second-order elastic tensor.

\section{Molecular dynamics simulation cell generation and statistics}
\begin{figure}[H]
\centering
\includegraphics[width=0.45\textwidth]{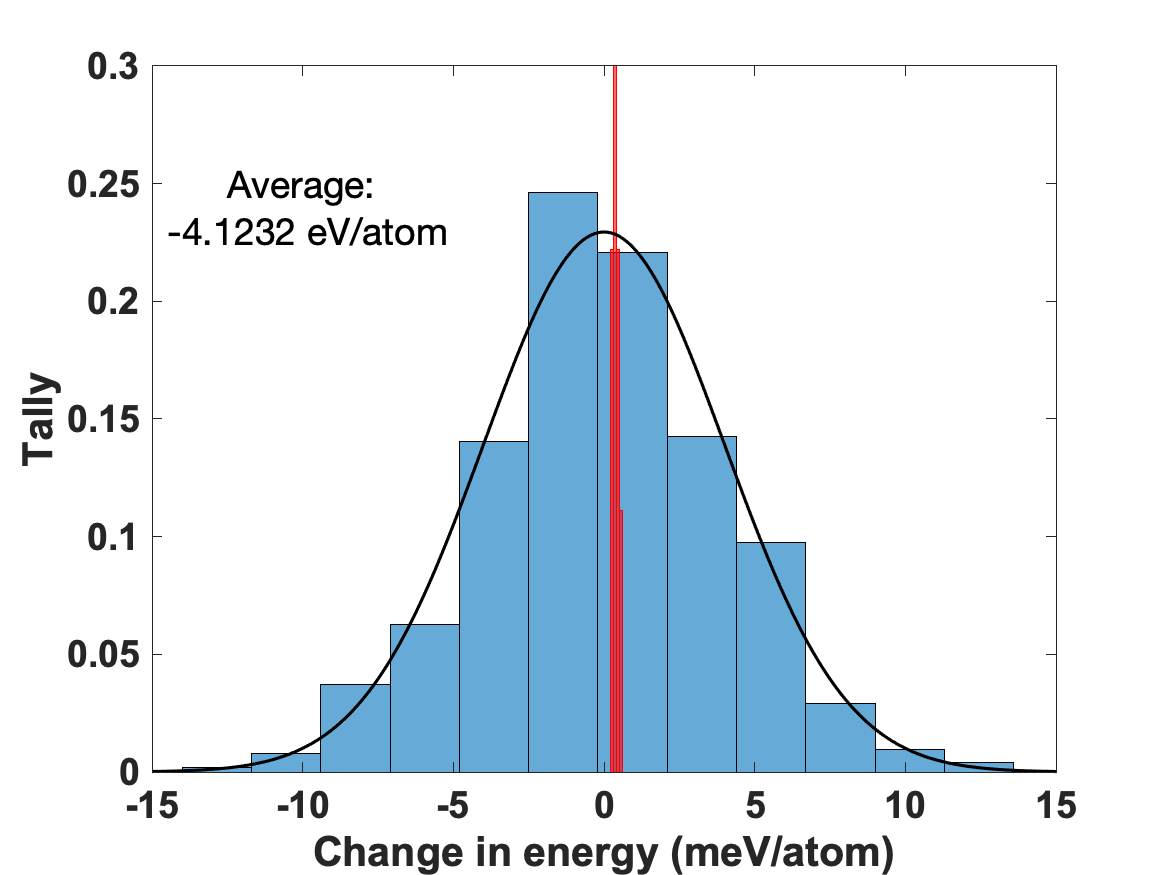}
\caption{The histogram shows the energy distribution per atom for 512 simulation cells containing $\sim500$ atoms random solid solution equiatomic CrFeCoNi (blue bars). A Gaussian distribution with the same standard deviation is illustrated with a solid black line. A histogram for nine larger cells containing $3,538,944$ atoms is portrayed with red bars. The average energy is very close for the large cells compared with the 512 sampling simulations. The average energy per atom is $\sim-4.1232$ eV/atom, and the distribution has been centered around this value for ease of visualization.}
\label{fig:MDHistogram}
\end{figure}

Fig. \ref{fig:MDHistogram} shows the distribution of energies per atom for 512 realization random generations of equiatomic CrFeCoNi with blue bars. The histogram has been shifted such that the average energy per atom ($\sim-4.1232$ eV/atom) corresponds to zero in the plot. The distribution shows a standard deviation of $\sim4$ meV/atom. A Gaussian function with the same deviation is shown for comparison purposes. The red bars correspond to nine realizations containing $3,538,944$ atoms. The average energy per atom is very close to the average of the 512 realizations. This comparison indicates that the number of local chemical environments in the large cells is properly sampled (since the energy converges to the statistical sampling using multiple random representations) and, therefore, supports our random generation supercell. 

\section{Electronic and phonon properties}
\begin{figure}[H]
\centering
\includegraphics[width=0.9\textwidth]{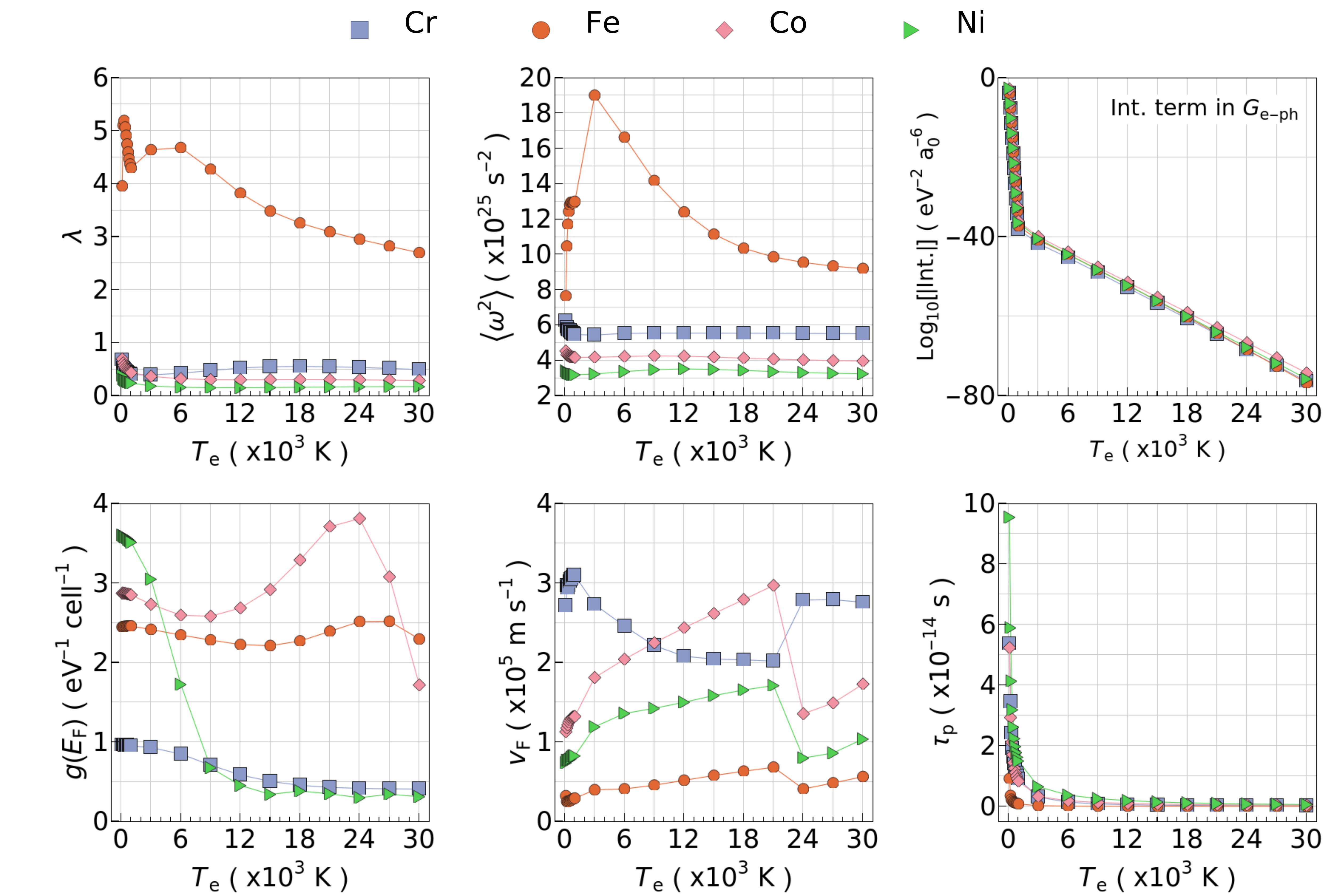}
\caption{Temperature dependence of the electronic and phonon quantities, necessary to calculate the electronic specific heat ($C_\mathrm{e}$), and the electronic thermal conductivity ($\kappa_\mathrm{e}$) of the base principal elements.}
\label{fig:PE-EPh-Prop-Comp}
\end{figure}

\begin{figure}[H]
\centering
\includegraphics[width=0.9\textwidth]{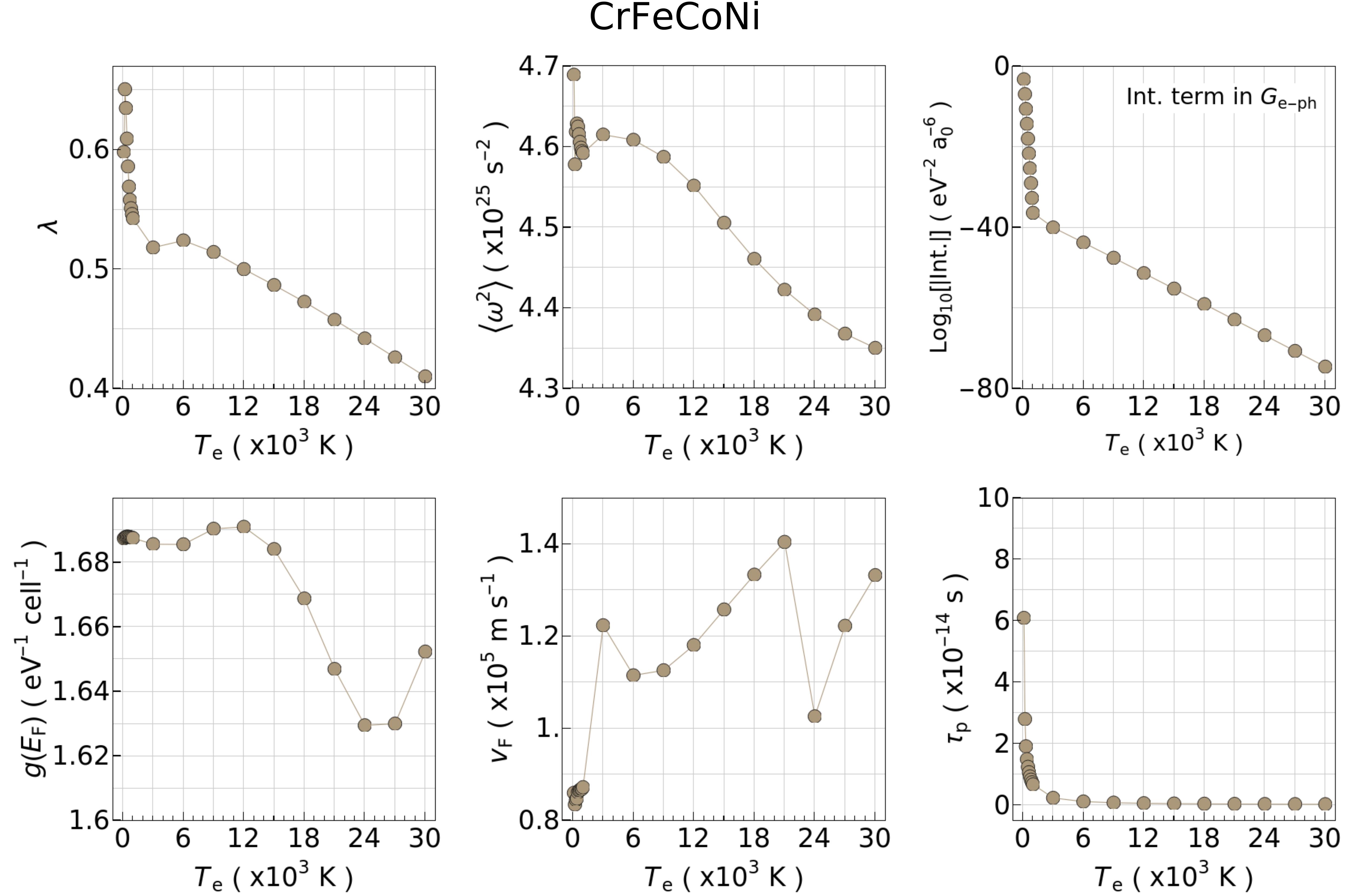}
\caption{Temperature dependence of the electronic and phonon quantities, necessary to calculate the electronic specific heat ($C_\mathrm{e}$), and the electronic thermal conductivity ($\kappa_\mathrm{e}$) of the face-centered cubic CrFeCoNi random solid solution.}
\label{fig:Xx4-EPh-Prop-Comp}
\end{figure}

\section{Composition dependence of electron and phonon properties}

\begin{figure}[H]
\centering
\includegraphics[width=0.9\textwidth]{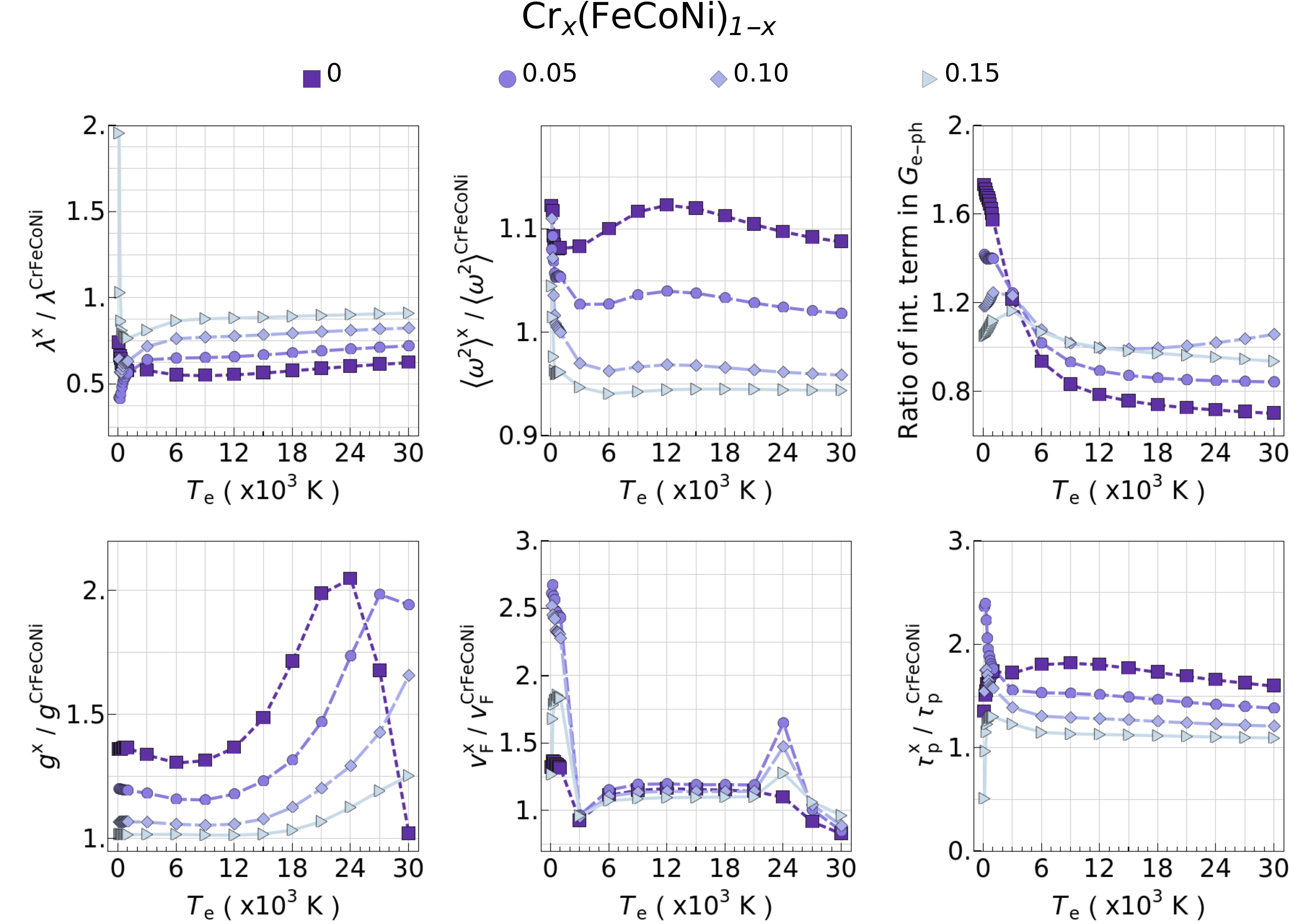}
\caption{Temperature dependence of the fractional electronic and phonon quantities, used in calculating the electronic specific heat ($C_\mathrm{e}$), and the electronic thermal conductivity ($\kappa_\mathrm{e}$),  of Cr$_x$(FeCoNi)$_{1-x}$  with respect to those of CrFeCoNi}. 

\label{fig:Cr-Xx3-EPh-Prop-Comp}
\end{figure}

\begin{figure}[H]
\centering
\includegraphics[width=0.9\textwidth]{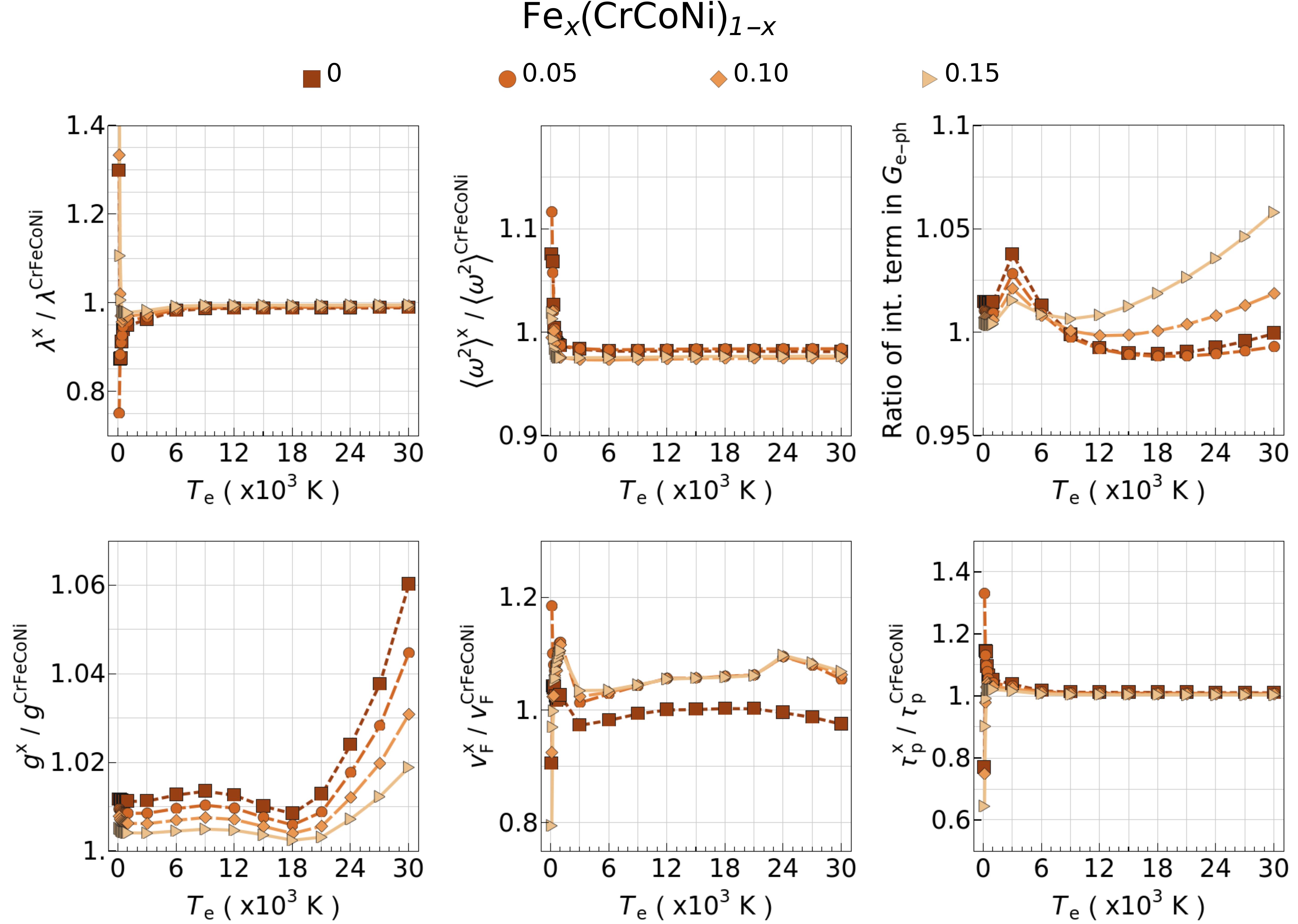}
\caption{Temperature dependence of the fractional electronic and phonon quantities, used in calculating the electronic specific heat ($C_\mathrm{e}$), and the electronic thermal conductivity ($\kappa_\mathrm{e}$),  of Fe$_x$(CrCoNi)$_{1-x}$  with respect to those of CrFeCoNi}. 
\label{fig:Fe-Xx3-EPh-Prop-Comp}
\end{figure}

\begin{figure}[H]
\centering
\includegraphics[width=0.9\textwidth]{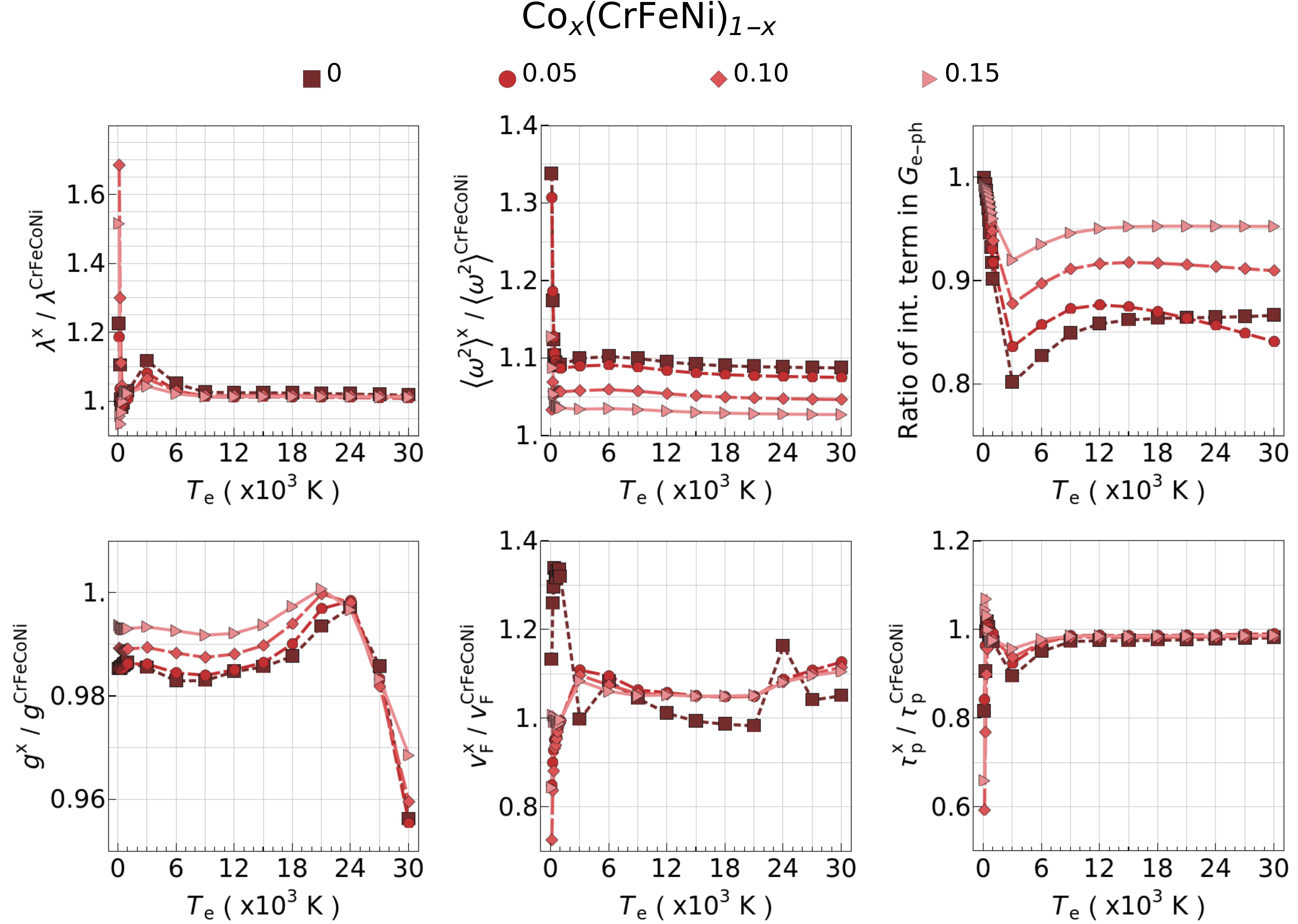}
\caption{Temperature dependence of the fractional electronic and phonon quantities, used in calculating the electronic specific heat ($C_\mathrm{e}$), and the electronic thermal conductivity ($\kappa_\mathrm{e}$),  of Co$_x$(CrFeNi)$_{1-x}$  with respect to those of CrFeCoNi}. 

\label{fig:Co-Xx3-EPh-Prop-Comp}
\end{figure}

\begin{figure}[H]
\centering
\includegraphics[width=0.9\textwidth]{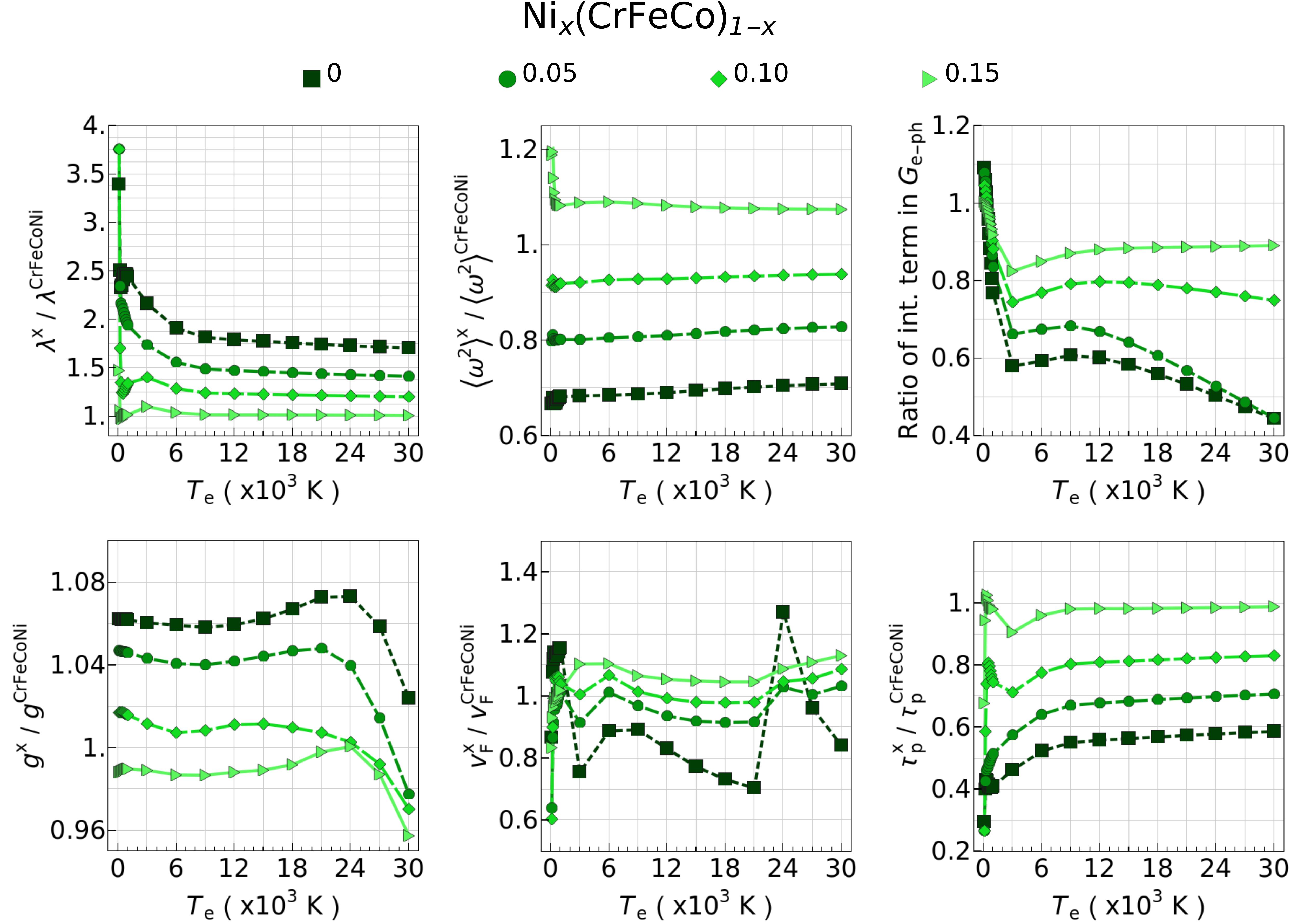}
\caption{Temperature dependence of the fractional electronic and phonon quantities, used in calculating the electronic specific heat ($C_\mathrm{e}$), and the electronic thermal conductivity ($\kappa_\mathrm{e}$),  of Ni$_x$(CrFeCo)$_{1-x}$  with respect to those of CrFeCoNi}. 
\label{fig:Ni-Xx3-EPh-Prop-Comp}
\end{figure}

\section{Equi-molar addition of Al, Mn, or Cu}

\begin{figure}[H]
\centering
\includegraphics[width=0.95\textwidth]{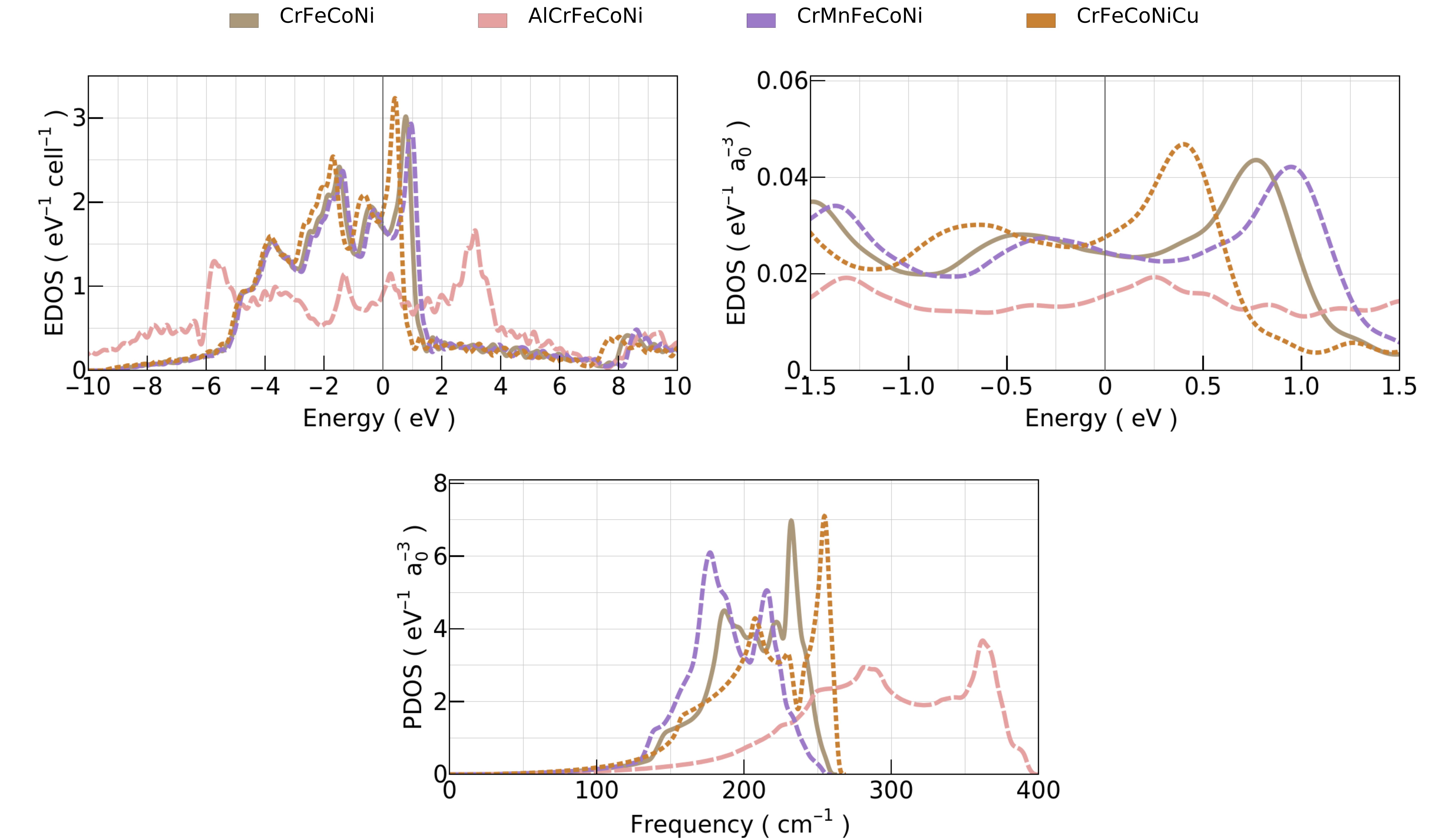}
\caption{Electronic density of states (EDOS), and phonon density of states (PDOS) of CrFeCoNi, AlCrFeCoNi, CrMnFeCoNi and CrFeCoNiCu.
Fermi level were set to zero in EDOS.} 
\label{fig:Xx5-EDOS-PDOS}
\end{figure}

\begin{figure}[H]
\centering
\includegraphics[width=0.9\textwidth]{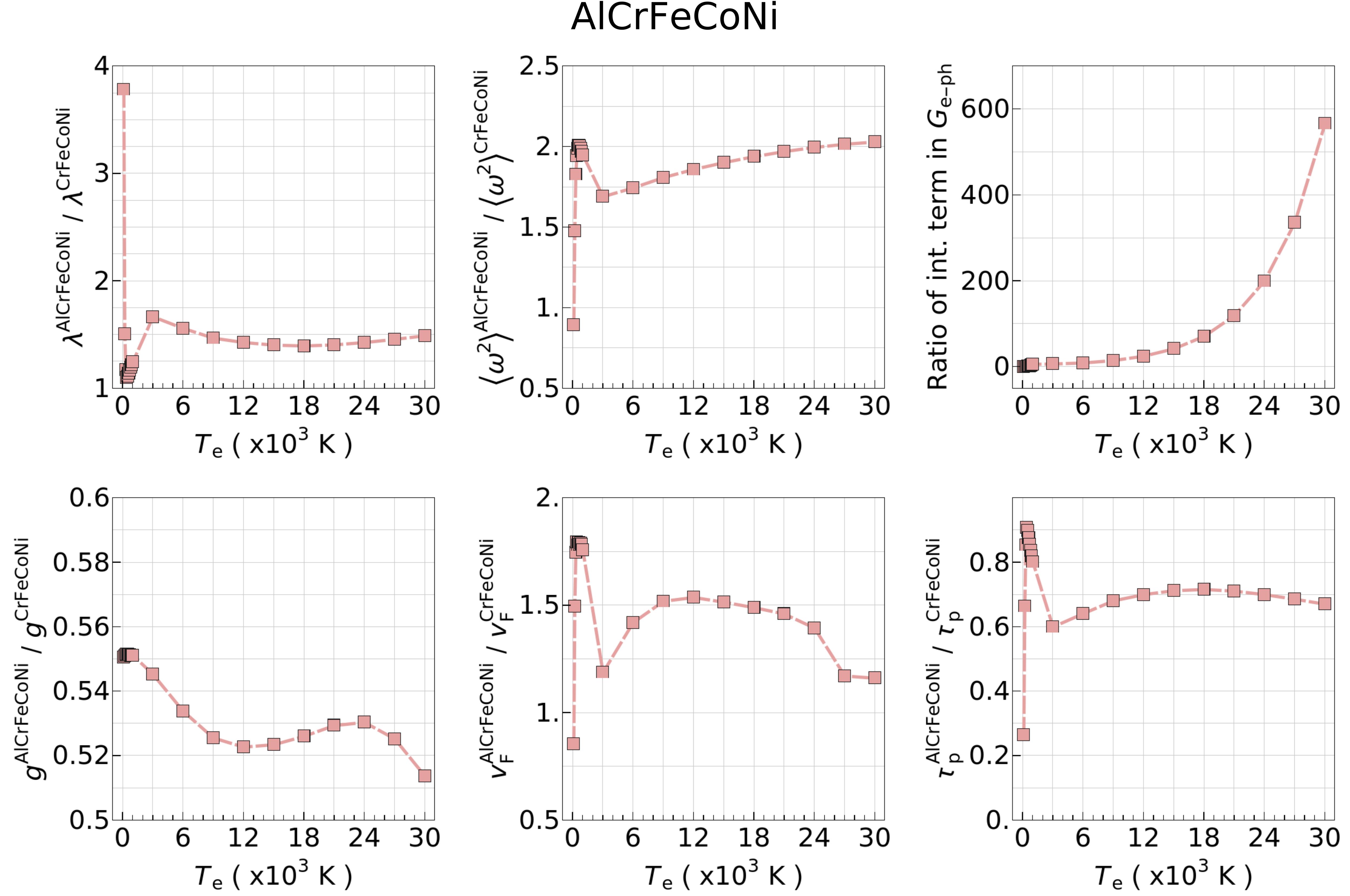}
\caption{Temperature dependence of the fractional electronic and phonon quantities, used in calculating the electronic specific heat ($C_\mathrm{e}$), and the electronic thermal conductivity ($\kappa_\mathrm{e}$),  of AlCrFeCoNi with respect to those of CrFeCoNi}. 
\label{fig:AlCrFeCoNi-EPh-Prop-Comp}
\end{figure}

\begin{figure}[H]
\centering
\includegraphics[width=0.9\textwidth]{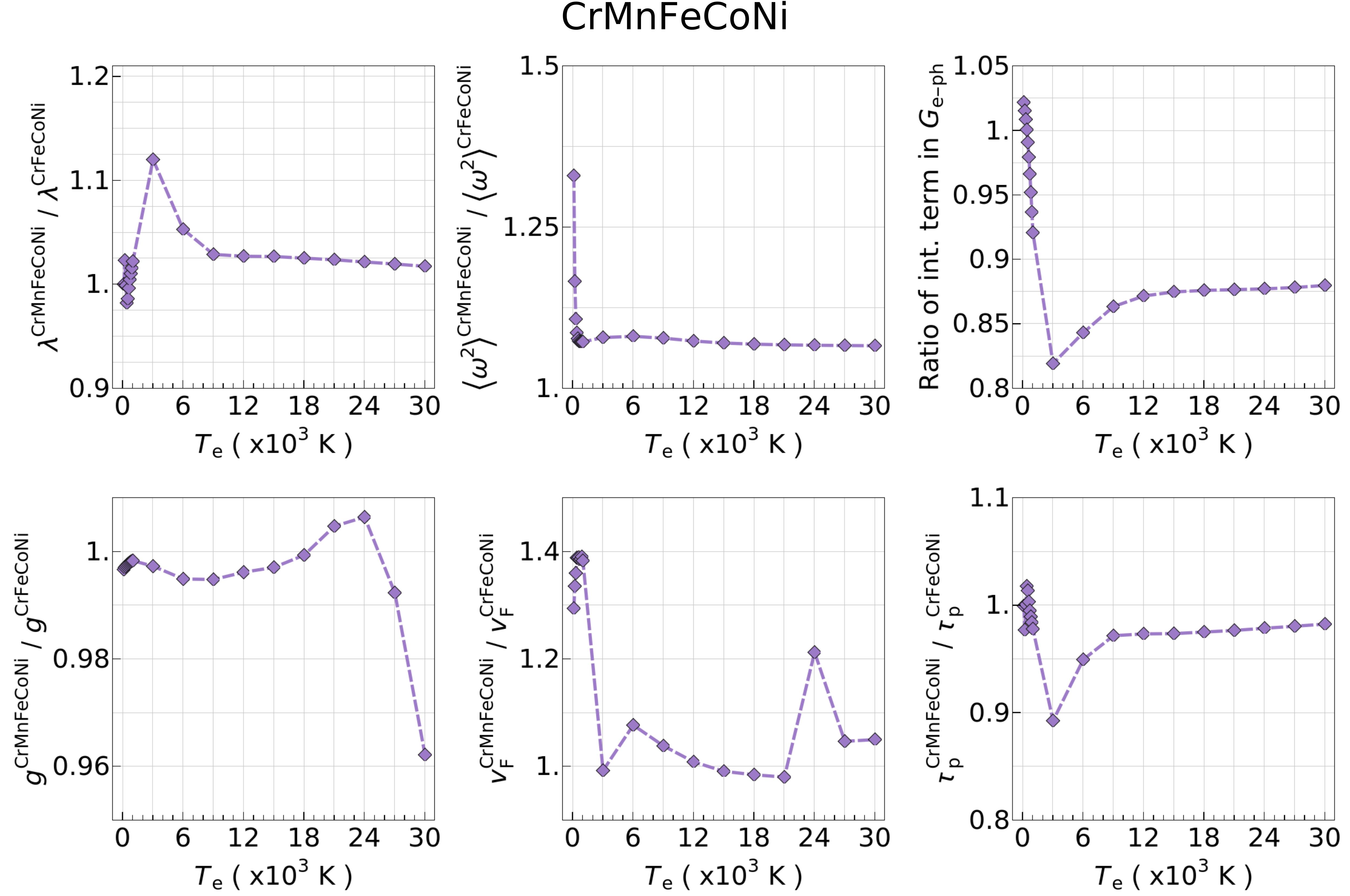}
\caption{Temperature dependence of the fractional electronic and phonon quantities, used in calculating the electronic specific heat ($C_\mathrm{e}$), and the electronic thermal conductivity ($\kappa_\mathrm{e}$),  of CrMnFeCoNi with respect to those of CrFeCoNi}. 
\label{fig:CrMnFeCoNi-EPh-Prop-Comp}
\end{figure}

\begin{figure}[H]
\centering
\includegraphics[width=0.5\textwidth]{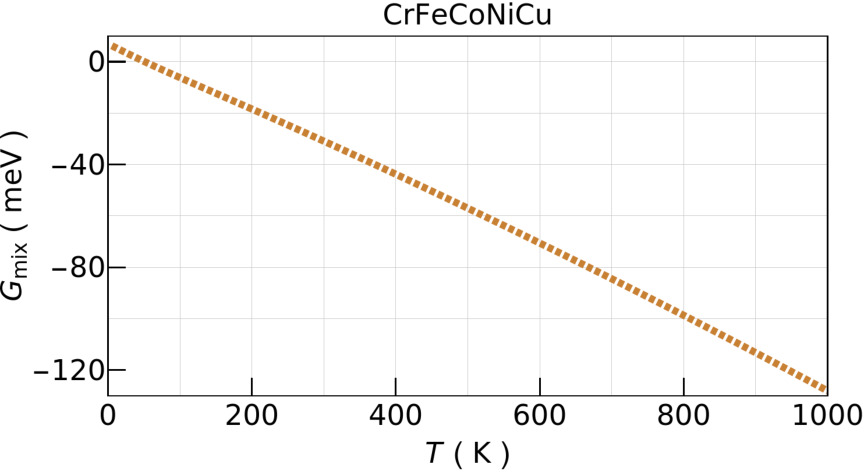}
\caption{Temperate dependence of the mixing Gibbs free energy ($G_\mathrm{mix}$) of the random solid solutions of CrFeCoNiCu.} 
\label{fig:CrMnFeCoNi-Gmix}
\end{figure}

\begin{figure}[H]
\centering
\includegraphics[width=0.9\textwidth]{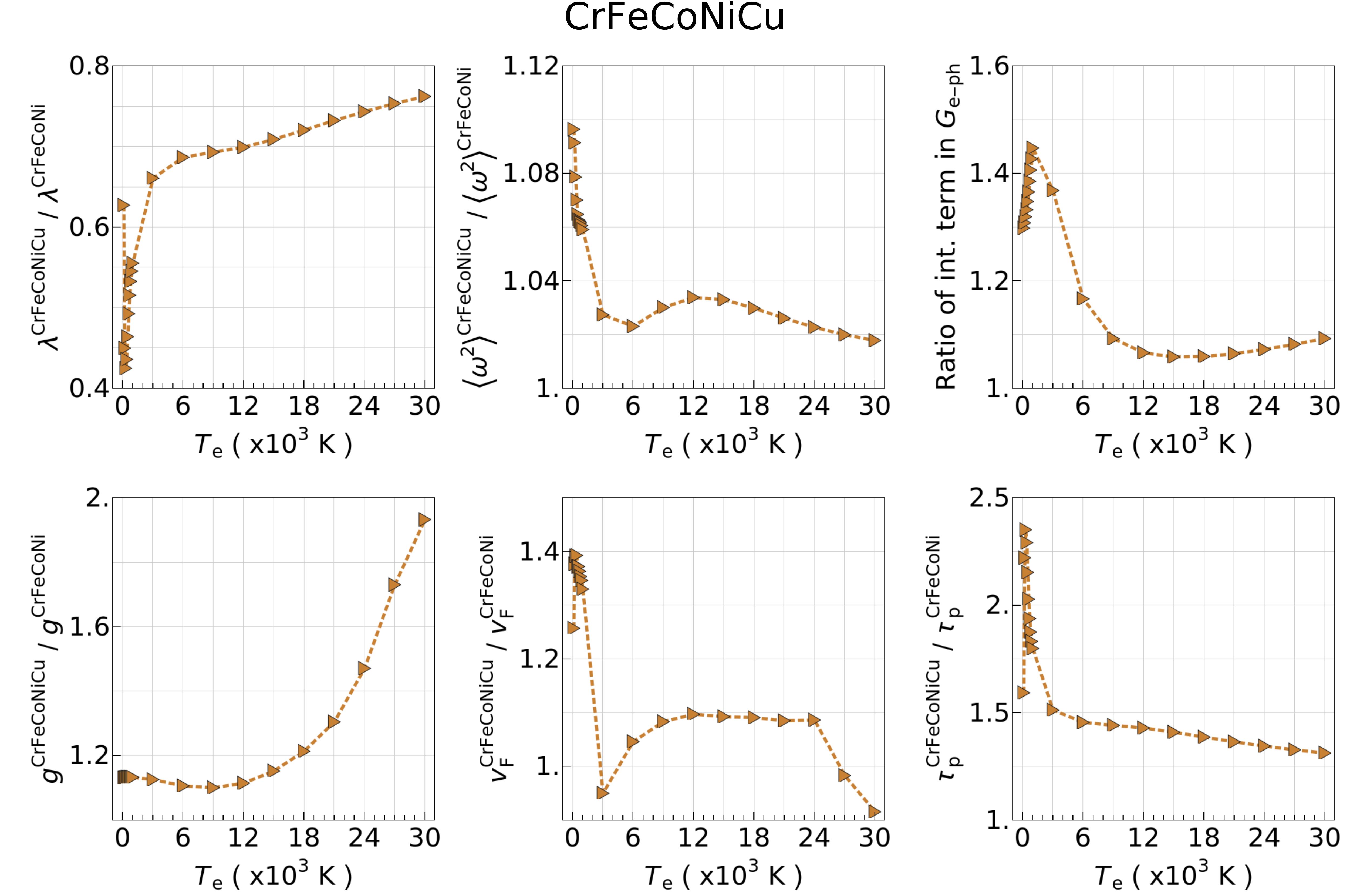}
\caption{Temperature dependence of the fractional electronic and phonon quantities, used in calculating the electronic specific heat ($C_\mathrm{e}$), and the electronic thermal conductivity ($\kappa_\mathrm{e}$),  of CrFeCoNiCu with respect to those of CrFeCoNi}. 
\label{fig:CrFeCoNiCu-EPh-Prop-Comp}
\end{figure}

\section{Convergence criteria for numbers of molecular dynamics simulations}

\begin{figure}[H]
\centering
\includegraphics[width=0.9\textwidth]{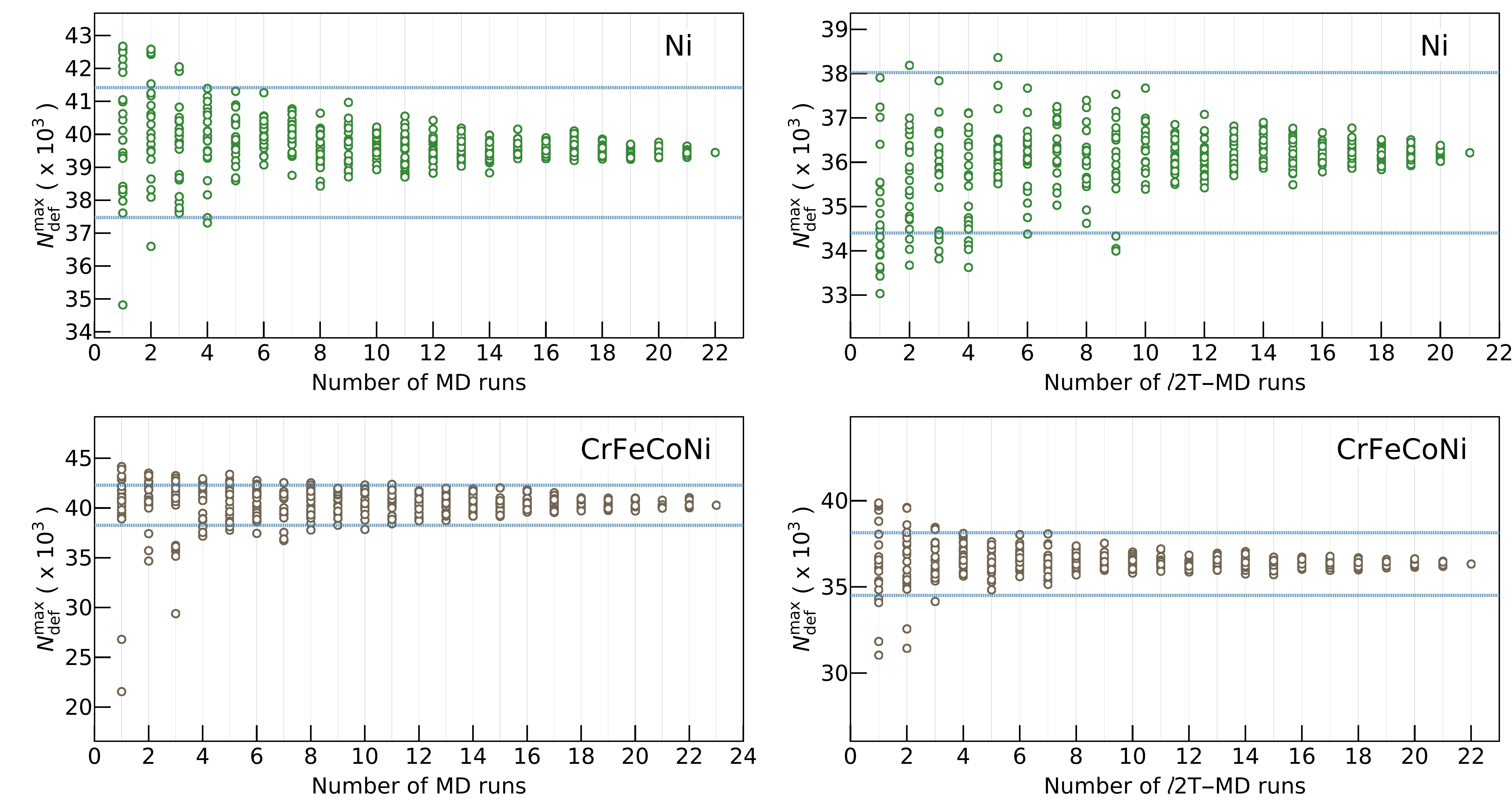}
\caption{The averaged maximum number of defected atoms at the peak of the thermal spike ($N_\mathrm{def}^\mathrm{max}$) for randomly selected subsets of MD and  $\ell$2T-MD simulations for a 50 keV PKA. 
Dashed blue lines show  $\pm 5 \%$ range of the averaged $N_\mathrm{def}^\mathrm{max}$ of all available simulations for each method and system.} 
\label{fig:Ndefmax-MDSimNum-Conv}
\end{figure}

\begin{figure}[H]
\centering
\includegraphics[width=0.9\textwidth]{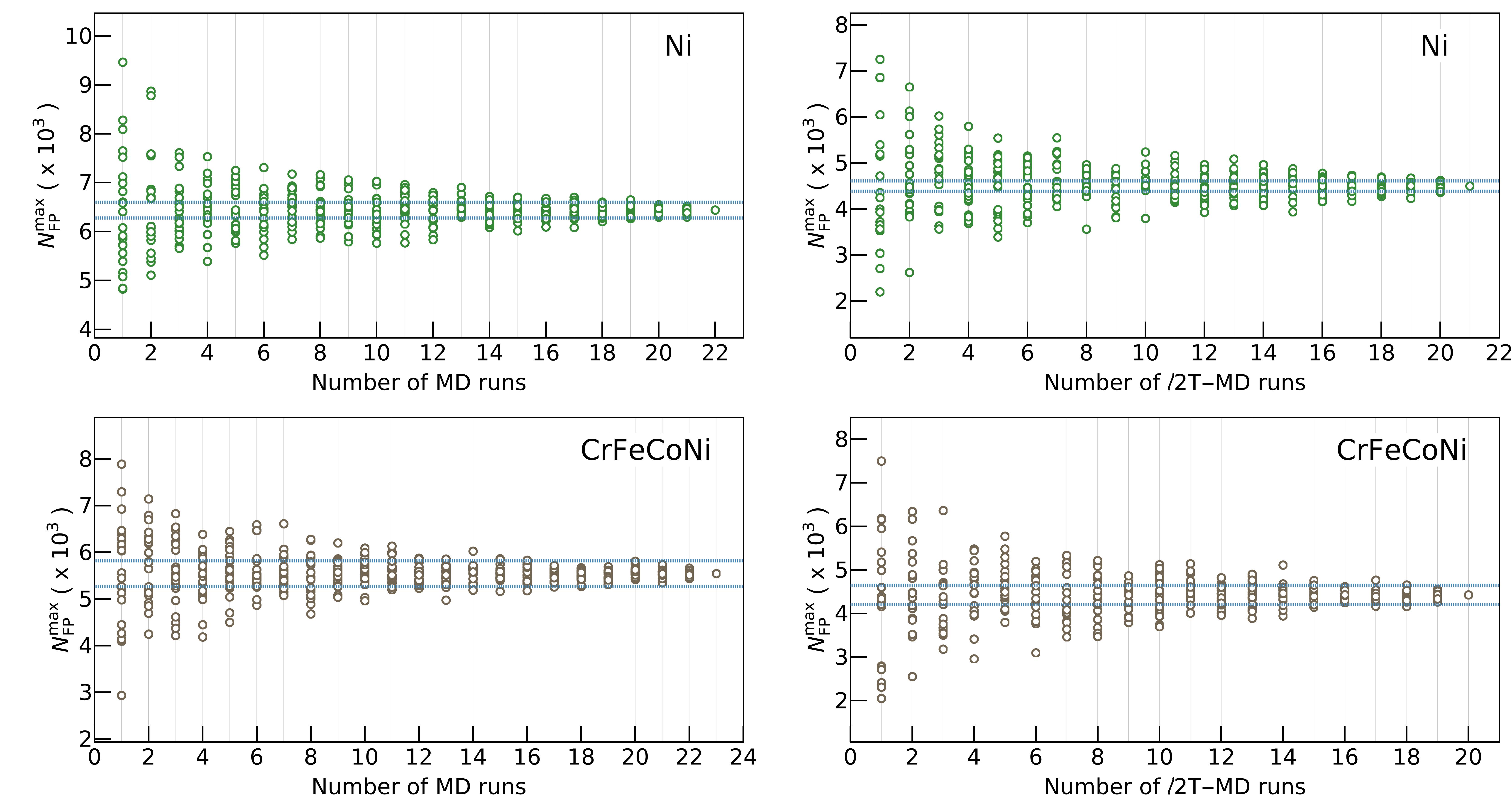}
\caption{The averaged maximum number of Frenkel pairs  ($N_\mathrm{FP}^\mathrm{max}$) for randomly selected subsets of MD and  $\ell$2T-MD simulations for a 50 keV PKA. 
Dashed blue lines show  $\pm 5 \%$ range of the averaged $N_\mathrm{FP}^\mathrm{max}$ of all available simulations for each method and system.} 
\label{fig:NFPmax-MDSimNum-Conv}
\end{figure}

\section{Assessment of the minimum cut-off energy for the electronic stopping}
\begin{table}[H]
\renewcommand{\arraystretch}{1.5} \setlength{\tabcolsep}{12pt}
\begin{center}
\begin{tabular}{llllll}
\hline \hline
   
      & $1$~eV &  $5$~eV & $10$~eV \\ \cline{2-4} 
$N_\mathrm{def}^\mathrm{max}$       & 36991 (299)  &  37831 (400) & 37934 (349) &  \\

$\tau_\mathrm{def}^\mathrm{max}$    & 0.31 (0.01) & 0.33 (0.01) & 0.33 (0.01) &  \\[0.25cm]

$N_\mathrm{def}^\mathrm{steady}$    &  2334 (75) & 2388 (91) & 2368 (90) &  \\

$\tau_\mathrm{def}^\mathrm{steady}$ & 15.52 (0.81) & 16.77 (0.69) & 16.60 (0.84) & \\[0.25cm]

$N_\mathrm{FP}^\mathrm{steady}$      &  100 (4) & 103 (4) & 100 (4) & \\

$\tau_\mathrm{FP}^\mathrm{steady}$   & 10.03	(1.07) & 9.83	(0.69) & 9.96 (0.95) & \\[0.25cm]

 \hline \hline
\end{tabular}
\end{center}
\caption{The mean values and standard error (in brackets) for the maximum number of defected atoms at the peak of the thermal spike ($N_\mathrm{def}^\mathrm{max}$), at steady state ($N_\mathrm{def}^\mathrm{steady}$), their corresponding times (in Picoseconds units) ($\tau_\mathrm{max}$, and $\tau_\mathrm{steady}$, respectively), and the number of Frenkel pairs
$N_\mathrm{FP}^\mathrm{steady}$ of CrFeCoNi within the conventional MD for the different minimum cut-off energy for the electronic stopping.}
\label{tab:Defect-Formation-ElStopping}
\end{table}

\section{Time evolution of defect formation in Ni, and CrFeCoNi}

\begin{table}[H]
\renewcommand{\arraystretch}{1.5} \setlength{\tabcolsep}{12pt}
\begin{center}
\begin{tabular}{lccccc}
\hline \hline
      &   \multicolumn{2}{c}{\underline{Ni}}    && \multicolumn{2}{c}{\underline{CrFeCoNi}}  \\ 
      & MD &  $\ell2$T-MD && MD &  $\ell2$T-MD \\ \cline{2-3} \cline{5-6} 

$\tau\; \text{at}\; T_\mathrm{lat}^\mathrm{dip}$ & 0.25 & 0.24 && 0.25 & 0.24 \\

$T_\mathrm{lat}^\mathrm{dip}$ & 344 & 340 && 346 & 340 \\

$T_\mathrm{el}\; \text{at}\; T_\mathrm{lat}^\mathrm{dip}$ &  - & 304 && - & 306 \\

$T_\mathrm{lat}^\mathrm{steady}$ & 355 & 334 && 356 & 332 \\

$T_\mathrm{el}^\mathrm{steady}$ & -  & 329 && - & 331 \\

 \hline \hline
\end{tabular}
\end{center}
\caption{Critical points of the averaged $T_\mathrm{avr}$, and $T_\mathrm{loc}^\mathrm{max}$ curves.}
\label{tab:SM-T-Curves}
\end{table}

\section{Self-interstitial atom and vacancy formation energies of the base elements}
Mono- and di-vacancy and self-interstitial atom (SIA) formation energy of each element in a CrFeCoNi simulation cells were calculated using molecular statics method in the Large-scale Atomic/Molecular Massively Parallel Simulator (LAMMPS) software~\cite{plimpton1993fast}. 
The simulation cells were subjected to energy minimization at $T=0$~K using  the Polak-Ribi\`ere conjugate gradient algorithm~\cite{10.1007/BFb0099521}. 
The calculations were repeated randomly for up to $1000$ different atomic sites to explore the statistical distribution of the formation energy of each element. 
The SIA sites studied are the $<100>$ dumbbell interstitial sites which was shown previously to be the most stable Self-interstitial with the lowest formation energy ~\cite{DELUIGI2021116951}. Point defect energies, including vacancy and SIA di-vacancy and di-SIA formation energies were calculated using the following equations 

\begin{align}\label{eq:seq8}
v_f^i=E_f(N-i)-\left[\frac{N_0-i}{N_0}\right]E_0,
\end{align}
\begin{align}\label{eq:seq9}
s_f^i=E_f(N+i)-\left[\frac{N_0+i}{N_0}\right]E_0,
\end{align}
where $v_f^i$ and $s_f^i$ are the point defect formation energies, $E_0$ is the cohesive energy per atom in the bulk, and $E_f(N-i)$ is the total energy in the simulation cell when the point defects are introduced. $N_0=702,464$ is the initial number of atoms in the simulation cells , $i$ is the number of added or removed atoms, which is $i=1$ for mono-vacancy formation energy $v_f^1$ and mono-SIA formation energy $s_f^1$ and $i=2$ for di-vacancy formation energy $v_f^2$ and di-SIA formation energy $s_f^2$.

\begin{figure}[H]
\centering
\includegraphics[width=0.95\textwidth]{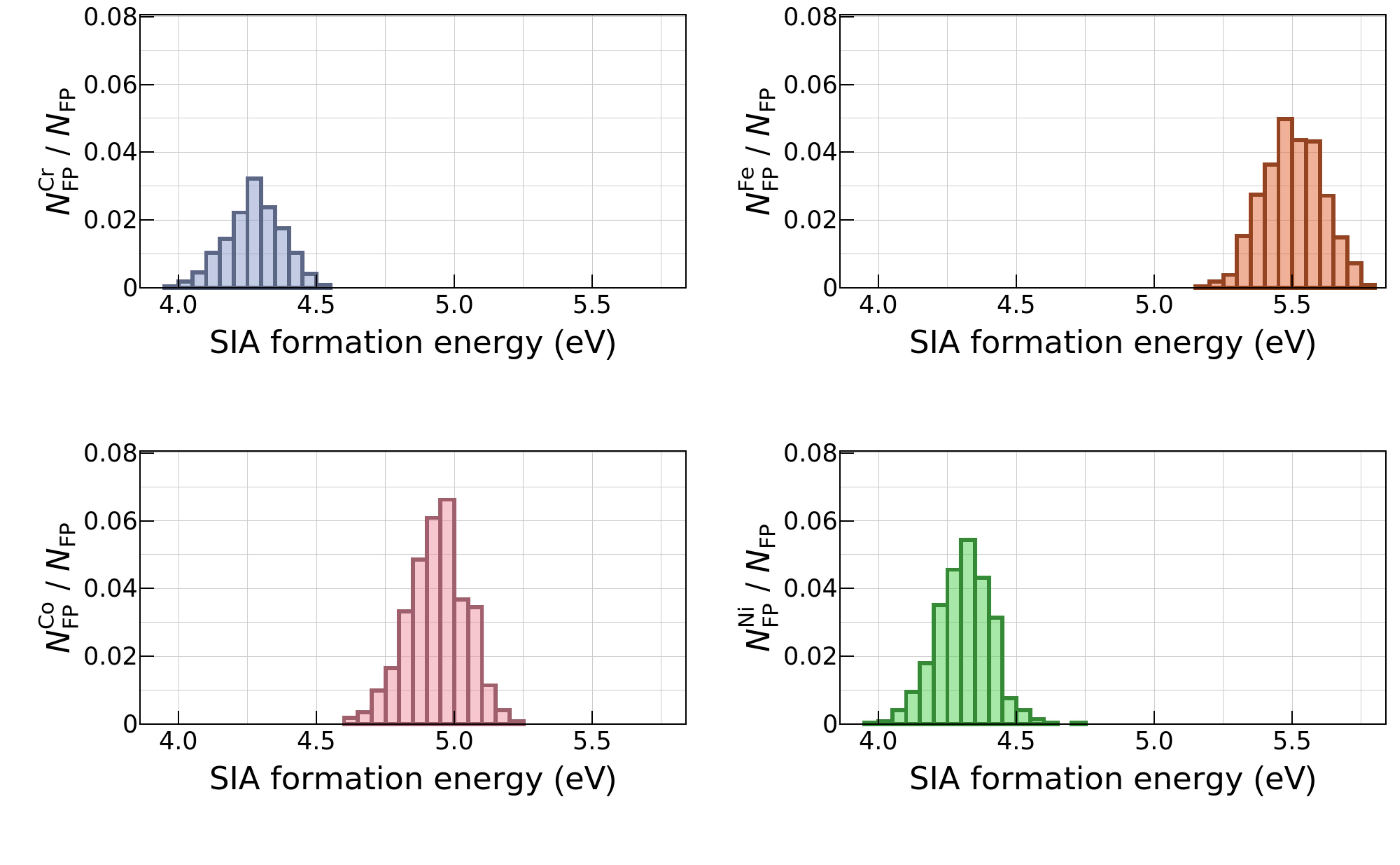}
\caption{Mono-self-interstitial atom (SIA) formation energies($s_f^1$) of the principal element of CrFeCoNi, calculated in CrFeCoNi environment. 
}
\label{fig:SIA-Formation-Energy-PE}
\end{figure}

\begin{figure}[H]
\centering
\includegraphics[width=0.95\textwidth]{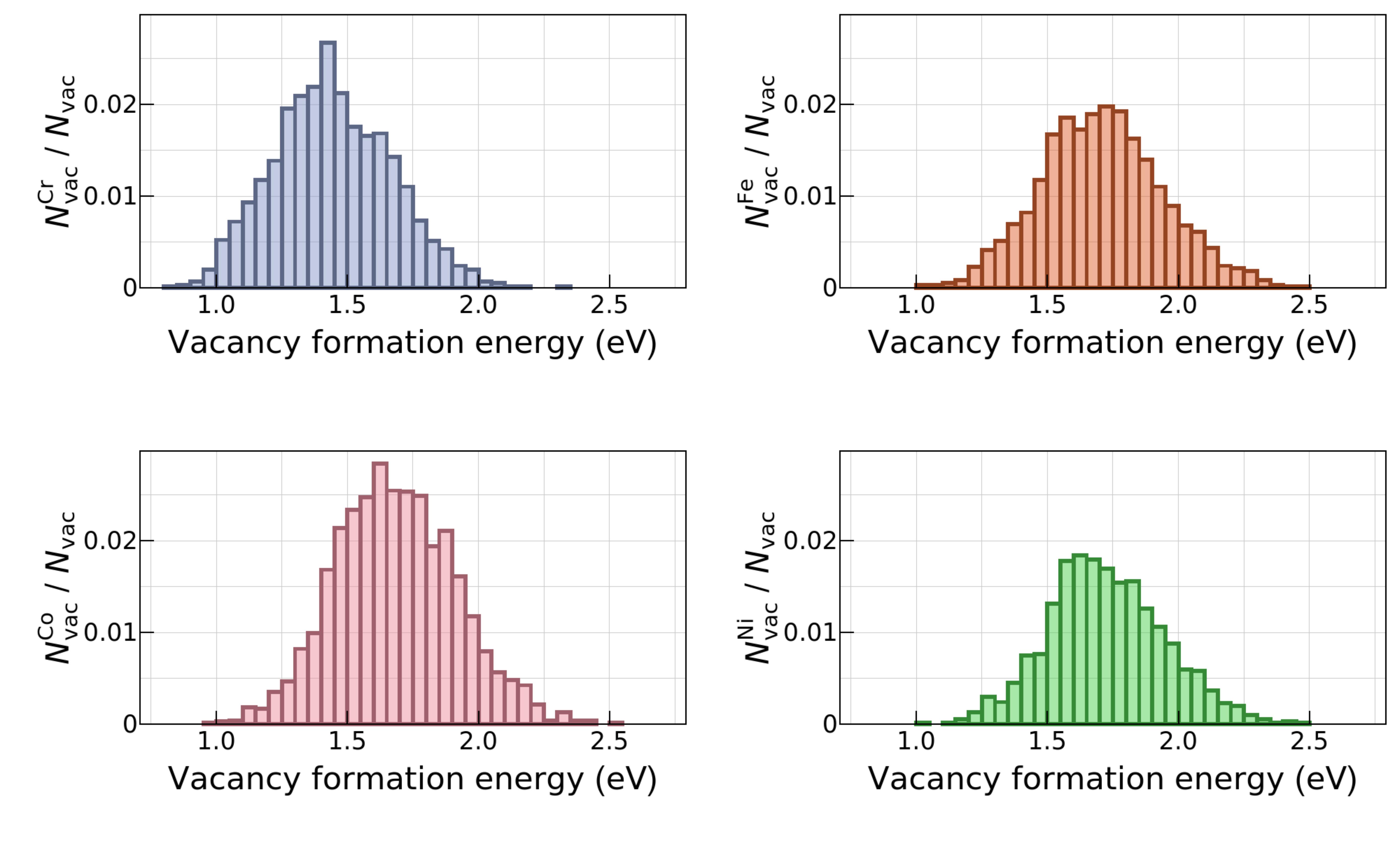}
\caption{Mono-vacancy formation energies ($v_f^1$) of the principal element of CrFeCoNi, calculated in CrFeCoNi environment. 
}
\label{fig:Vac-Formation-Energy-PE}
\end{figure}

\begin{table}[H]
\renewcommand{\arraystretch}{1.5} \setlength{\tabcolsep}{12pt}
\begin{center}
\begin{tabular}{llllll}
\hline \hline
      & $v_f^1$ &  $v_f^2$ && $s_f^1$ & $s_f^2$ \\ \cline{2-3} \cline{5-6} 
Cr & 1.46 (0.23)  &  2.99 (0.30) & & 4.27 (0.10)  & 13.38 (0.62) \\

Fe  & 1.71 (0.23) & 3.38 (0.33) & & 5.51 (0.11) & 15.65 (1.25) \\

Co &  1.69 (0.23) & 3.24 (0.32) & & 4.94 (0.10) & 14.95 (0.78) \\

Ni  & 1.70 (0.22) & 3.31 (0.31) & & 4.31 (0.10) & 13.19 (0.67) \\[0.25cm]

 \hline \hline
\end{tabular}
\end{center}
\caption{The mean values and standard deviation (in brackets) for $v_f^1$,  $v_f^2$, $s_f^1$ and $s_f^2$.}
\label{tab:Defect-Formation-SM}
\end{table}

\section{Defect clustering}

\begin{figure}[H]
\centering
\includegraphics[width=0.45\textwidth]{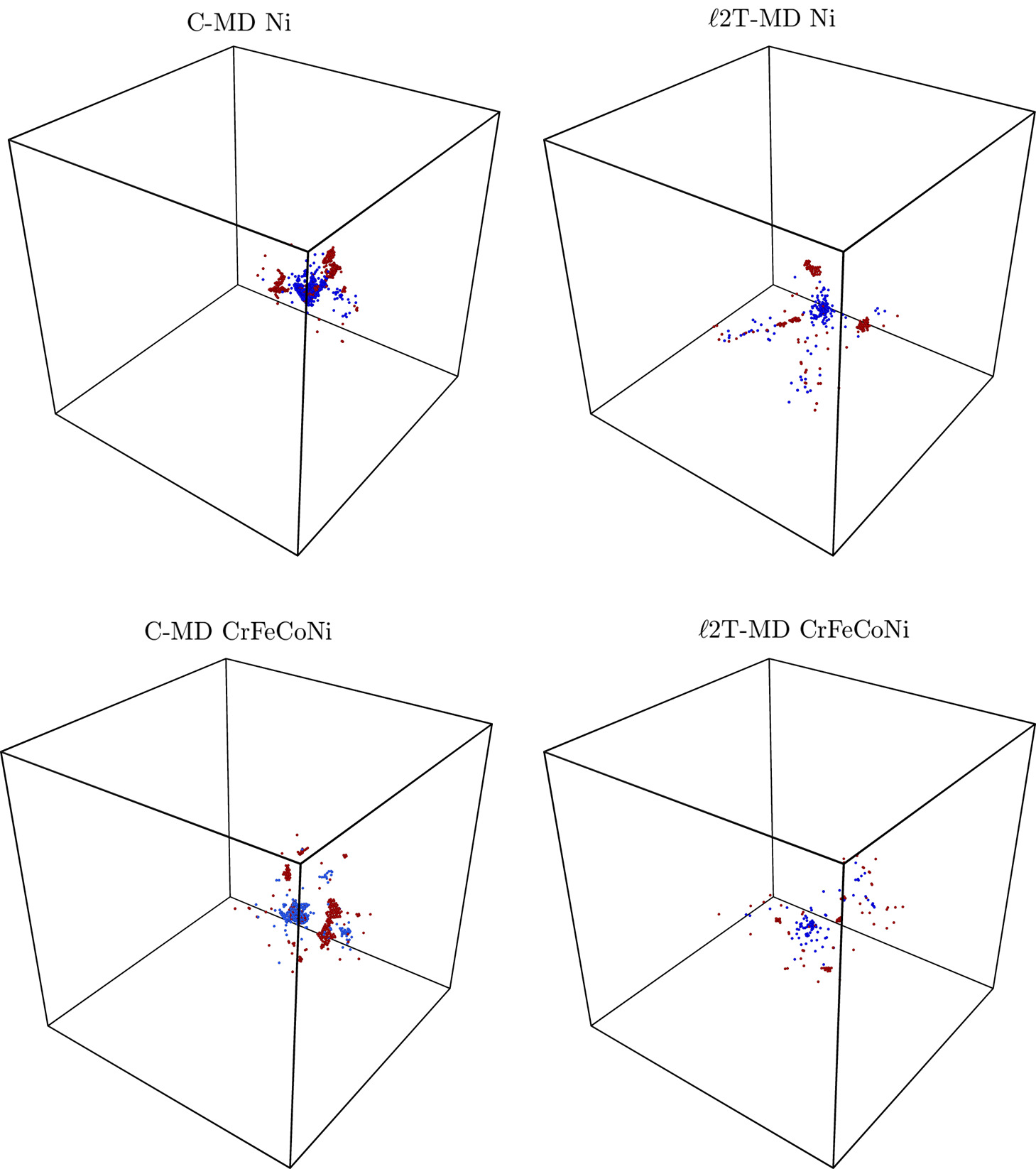}
\caption{Self-interstitial atom (in red), and vacancy (in blue) populations at the steady states of Ni, and CrFeCoNi within the conventional MD and the $\ell$2T-MD simulations.}
\label{fig:FP-Clustering}
\end{figure}